%% file: main.tex
\documentclass[pdflatex, sn-nature]{sn-jnl}
\usepackage{graphicx}%
\usepackage{multirow}%
\usepackage{amsmath,amssymb,amsfonts}%
\usepackage{newtxtext}
\usepackage{amsthm}%
\usepackage{mathrsfs}%
\usepackage{xcolor}%
\usepackage{textcomp}%
\usepackage{manyfoot}%
\usepackage{booktabs}%
\usepackage{algorithm}%
\usepackage{algorithmicx}%
\usepackage{algpseudocode}%
\usepackage{listings}%
\usepackage{makecell}
\usepackage{hhline}
\usepackage{tabularx}
\usepackage{array}
\usepackage{csquotes}
\usepackage{caption}
\usepackage{comment}
\usepackage{longtable}
\usepackage{siunitx}
\usepackage{ltxtable}
\usepackage{float}
\usepackage{mathtools}
\usepackage{bbm}
\usepackage{hyperref}
\usepackage{lineno}

\usepackage[T1]{fontenc}
\usepackage{fancyvrb}

\DefineVerbatimEnvironment{code}{Verbatim}{
  fontsize=\small,
  codes={\catcode`\~=13 \def~{\raisebox{-0.35ex}{\textasciitilde}}}
}

\begin{document}


\title{Perceived face similarity reflects inverse-generative and naturalistic discriminative objectives}




\author*[1]{\fnm{Wenxuan} \sur{Guo}}\email{w.guo@columbia.edu}

\author[2]{\fnm{Heiko H.} \sur{Schütt}}\email{heiko.schutt@uni.lu}

\author[3]{\fnm{Kamila Maria} \sur{Jozwik}}\email{kj287@cam.ac.uk}

\author[4]{\fnm{Katherine R.} \sur{Storrs}}\email{katherine.storrs@auckland.ac.nz}

\author*[1,5]{\fnm{Nikolaus} \sur{Kriegeskorte}}\email{n.kriegeskorte@columbia.edu}
\equalcont{These authors jointly supervised this work.}

\author*[6,7]{\fnm{Tal} \sur{Golan}}\email{golan.neuro@bgu.ac.il}
\equalcont{These authors jointly supervised this work.}

\affil[1]{\orgdiv{Department of Psychology}, \orgname{Columbia University},  \orgaddress{\city{New York}, 
\state{NY}, \country{USA}}}

\affil[2]{\orgdiv{Department of Behavioural and Cognitive Sciences}, \orgname{Université du Luxembourg}, \orgaddress{\city{Esch-sur-Alzette}, \country{Luxembourg}}}

\affil[3]{\orgdiv{MRC Cognition and Brain Sciences Unit}, \orgname{University of Cambridge}, \orgaddress{\city{Cambridge}, 
\country{England}}}

\affil[4]{\orgdiv{School of Psychology}, \orgname{University of Auckland}, \orgaddress{\city{Auckland}, \country{New Zealand}}}

\affil[5]{\orgdiv{Department of Neuroscience}, \orgname{Columbia University}, \orgaddress{\city{New York}, \state{NY}, \country{USA}}}

\affil[6]{\orgdiv{Department of Industrial Engineering and Management}, \orgname{Ben-Gurion University of the Negev}, \orgaddress{\city{Be’er Sheva}, \country{Israel}}}

\affil[7]{\orgdiv{School of Brain Sciences and Cognition}, \orgname{Ben-Gurion University of the Negev}, \orgaddress{\city{Be’er Sheva}, \country{Israel}}}

\keywords{face perception, deep neural networks, face similarity judgments, controversial stimuli}

\abstract{The perceptual representations supporting our ability to recognize faces remain a computational mystery. Deep neural networks offer mechanistic hypotheses for human face perception, but theoretically distinct models often make indistinguishable representational predictions for randomly sampled faces. To expose diagnostic differences among these hypotheses, we compared six neural network models sharing an architecture but trained on distinct tasks, using face pairs optimized to elicit contrasting model predictions (``controversial'' pairs) alongside randomly sampled pairs. We tested model predictions against face-dissimilarity judgments from 864 human participants across stimulus sets differing in realism and pose variation. Models prioritizing high-level, invariant structures (trained via inverse rendering, face identification, or object classification) most robustly matched human judgments. Furthermore, models trained on natural images typically outperformed synthetic-trained counterparts. Together, these findings suggest that human face perception is shaped by mechanisms that infer latent causes of facial appearance, discount nuisance variation, and are tuned by natural image statistics.}
\maketitle

\section{Introduction}\label{sec1}
Deep neural network models (NNs) enable us to implement and test computational theories of face perception \citep{otoole_face_2021, dyck_modeling_2023, yildirim_efficient_2020, ryali_leveraging_2020, daube_grounding_2021, BLAUCH2021104341, peterson_deep_2022, jozwik_face_2022, dobs_behavioral_2023, jiahui_modeling_2023,shoham_visual_semantic_2024}. In addition to achieving human-level face recognition performance \citep[e.g.][]{taigman_deepface_2014, parkhi_deep_2015, sun_deeply_2015, lu_surpassing_2015, Schroff_2015, liu_sphereface_2017, Deng_2019_CVPR}, these models provide explicit, image-computable hypotheses about the representations and computations underlying face perception. A growing body of work shows that NN representations quantitatively predict human behavior in face perception tasks across diverse experimental designs, capturing not only robust perceptual regularities but also signature modes of failure \citep{peterson_deep_2022, jozwik_face_2022, dobs_behavioral_2023, jiahui_modeling_2023, yildirim_efficient_2020, tian_face_2022, xu_face_2021, yovel_deep_2023}. These advances make NNs increasingly plausible as a high-level mechanistic account of the computations that underlie human face perception.

Despite this progress, the field is far from converging on a single computational model of face perception. One key question is what objectives govern the formation of face representations. Candidate objectives include face identification \cite{abudarham_critical_2019, BLAUCH2021104341, jacob_qualitative_2021, tian_face_2022, dobs_behavioral_2023, rosemblaum_concurrent_2025}, domain-general object recognition \cite{gauthier_behrmann_1999, grossman_convergent_2019, vinken2023, chang_explaining_2021}, inverse-generative objectives that infer latent causes of images \cite{yildirim_efficient_2020, ryali_leveraging_2020}, and generative objectives that support learning the distribution of faces via latent-variable modeling \cite{daube_grounding_2021}. If optimization of an NN for a particular objective gives rise to representations that match human perception, this helps us understand \textit{why} the representations take this form \citep{kanwisher2023240}.

To date, research comparing neural network representations to human face perception has not yet yielded conclusive evidence of the underlying objective. A remarkable diversity of models achieves comparable performance in predicting human similarity judgments when evaluated on randomly sampled sets of face images \cite{jozwik_face_2022}. The indistinguishable models include image-computable models such as NNs and non-image-computable models such as ground-truth generative latent variables. Despite implementing very different theories of face representation, both classes of model approach the human behavioral noise ceiling (the maximum performance a model can achieve given the intra- and inter-subject variability). A closely analogous pattern is observed when predicting cortical representational geometries for randomly sampled sets of face stimuli \cite{carlin_adjudicating_2017}. This ``fortunate conundrum''---that multiple qualitatively distinct models perform similarly well---also arises in studies of the representation of general (non-face) natural images: when evaluated on broad random samples of natural images, qualitatively distinct NNs often achieve indistinguishable predictive accuracy \cite{jozwik_face_2022, storrs_2021_diverse, carlin_adjudicating_2017, grossman_convergent_2019, conwell_large-scale_2024, schrimpf_integrative_2020, golan_controversial_2020}.

Why do distinct computational models achieve similar prediction performance for randomly selected face images? Two factors may contribute: (1) NN models are highly flexible function approximators, and different models may converge in their predictions within the distribution that the training stimuli were drawn from. (2) Different levels of dissimilarity may be confounded: pairs of faces that differ more in high-level or holistic properties such as overall aspect ratio, 3D shape, facial expression, gender, or race, also differ more in terms of low-level properties, such as color, local curvature, and surface texture.

\begin{figure}[H]
\centering
\makebox[\linewidth][c]{\includegraphics[scale=0.75]{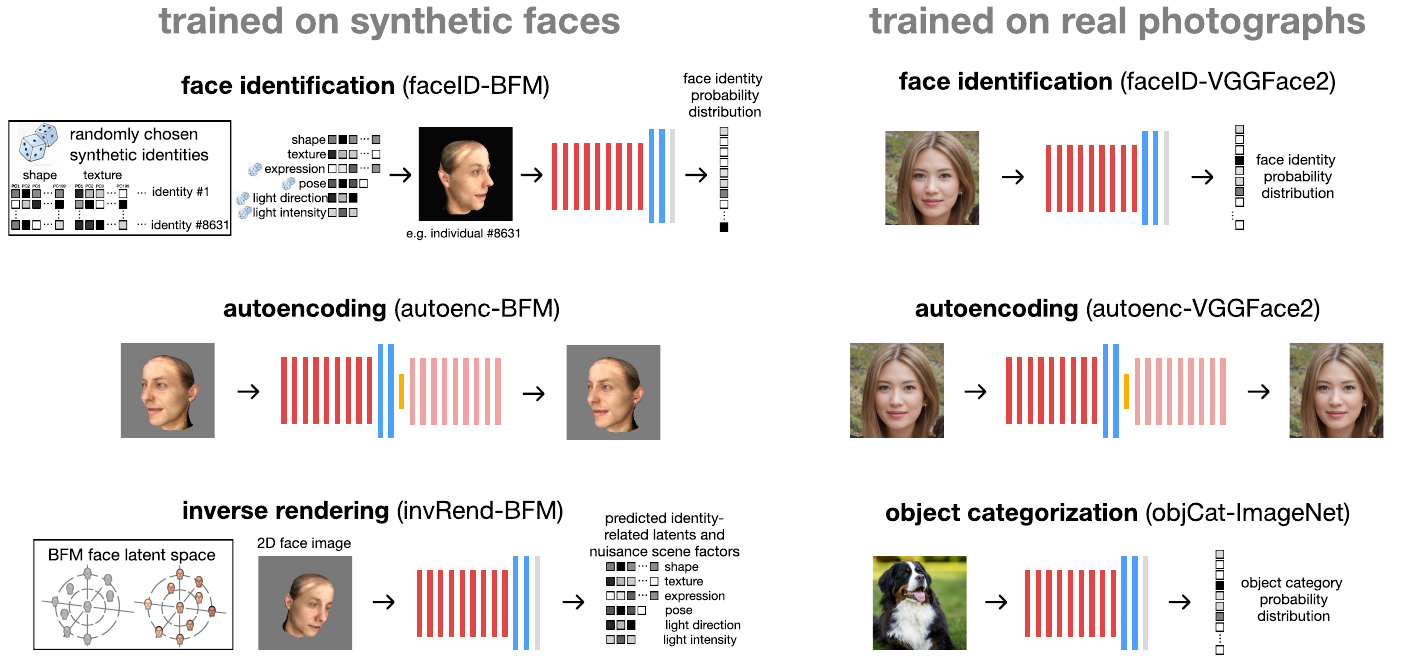}}
\caption{\textbf{Six candidate models of human face dissimilarity judgments}. All models used the same VGG-16 convolutional neural network (CNN) architecture \citep{simonyan_vgg16_2015} and varied along two factors: training objective (discriminative vs. generative) and visual diet (synthetic, left column, vs. real photographs, right column).
The face-identification models (first row) were trained to map face images to identity labels: in faceID-BFM, identities were defined by BFM shape and texture latents, while expression, pose, light direction, and light intensity were sampled as nuisance variables to generate multiple images of each synthetic identity; in faceID-VGGFace2, identities were defined by person labels in VGGFace2 \citep{cao_vggface2_2018}. 
autoenc-BFM and autoenc-VGGFace2 (second row) are variational autoencoders trained to reconstruct the input image while regularizing the latent representation toward a Gaussian prior \citep{kingma_auto-encoding_2014}; we used the SigmaVAE formulation \citep{rybkin_sigmavae_2020}. 
The inverse-rendering model, invRend-BFM, was trained to explicitly infer the BFM generative parameters of a 2D face image, including identity-related shape and texture latents as well as expression, pose, light direction, and light intensity. 
The object-categorization model, objCat-ImageNet, was trained to classify natural images into ImageNet object categories \citep{deng_imagenet_2009}. 
Details of the training objective, architectural modifications, and training dataset for each model are provided in Methods~\ref{sec:model_training}. 
For copyright reasons, the VGGFace2 \cite{cao_vggface2_2018} images in this figure were replaced with synthetic faces generated using \url{https://thispersondoesnotexist.com/}; the ImageNet \cite{deng_imagenet_2009} illustration was obtained from \url{https://commons.wikimedia.org/}.}
\label{fig:six_models}
\end{figure}

The challenge, then, is to design face stimuli for which different models predict distinct experimental outcomes and in which higher-level (e.g., identity-related) features are not systematically entangled with low-level (pixel-space) features. We manipulate three aspects of stimulus-set design: (1) whether faces are randomly sampled or optimized to be controversial among the candidate perceptual models \cite{golan_controversial_2020,golan22,golan_testing_2023}, (2) whether pose is held constant or allowed to vary, allowing us to dissociate higher-level facial features from lower-level image properties, and (3) whether faces are generated by the Basel Face Model (BFM; \cite{blanz_morphable_1999, bfm09, bfm17}), which affords greater experimental control, or by StyleGAN3 \citep{karras_alias-free_2021}, which affords greater photorealism. The resulting stimulus sets are then evaluated in a large-scale behavioral experiment, in which participants arrange face pairs according to the degree of perceived similarity within each pair \citep{valentine_unified_1991, kriegeskorte_inverse_2012}. We then used these stimulus sets to evaluate six computational hypotheses (\figurename~\ref{fig:six_models}) in terms of their ability to explain human face dissimilarity judgments.

Together, these computational models and behavioral experiments enable us to address three fundamental questions about human face perception: First, what computational objectives best explain human face-dissimilarity judgments? The answer contributes to a high-level understanding of human face perception because it tells us what the system is optimized for. Second, what visual diet for model training yields human-like face representations? The answer reveals to what extent different aspects of natural visual experience with faces shape human face perception. Finally, which models best explain human dissimilarity judgments under face-pose variation? The answer constrains computational theories of how robustness to face pose is achieved in the human visual cortex.

\section{Results}\label{sec2}
We tested six candidate models in three experiments using different stimulus families, each providing a different balance of experimental control and ecological realism: frontal BFM faces, pose-varied BFM faces, and photorealistic faces generated by StyleGAN3. For each stimulus family, we generated 12 stimulus sets using random sampling and 12 stimulus sets using controversial stimulus synthesis, in which sets of face pairs were optimized to be controversial among the models in terms of their predicted dissimilarities (see Methods~\ref{sec:controversial_face_synthesis}). Each stimulus set consisted of 72 face pairs grouped into 12 trials of six pairs and was judged by a separate group of 12 randomly allocated participants.

\begin{figure}[t]
\centering
\makebox[\linewidth][c]{\includegraphics[scale=1.2]{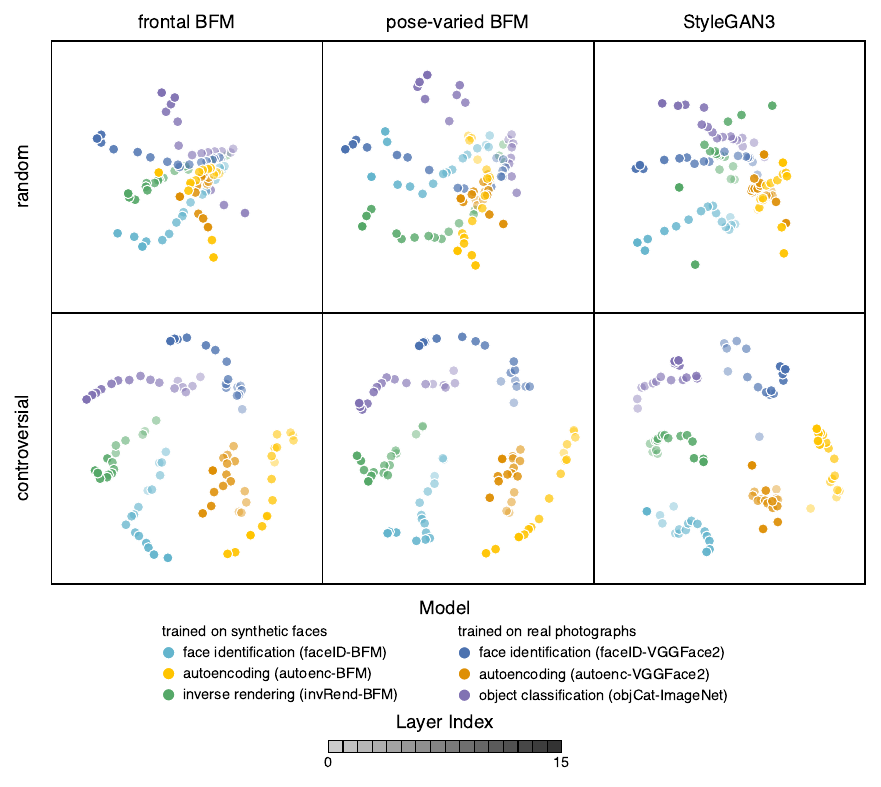}}
\caption{\textbf{Controversial sampling separates model representational geometries}. Multidimensional scaling (MDS) visualization of pairwise similarities between all model layers ($6$ models $\times 16$ layers) under random sampling (top row) and controversial sampling (bottom row). Each point represents a single model layer. Models are distinguished by color, and layer index is indicated by transparency (earlier layers are more transparent). Layer-to-layer similarity was computed as Spearman's correlation between layer-wise dissimilarity predictions across face pairs within trials, averaged across trials and seeds. We then fit a two-dimensional metric MDS embedding to the corresponding precomputed layer-to-layer dissimilarity matrix, minimizing unweighted raw stress.
Under random sampling, early layers from different models are close in representational space. Controversial sampling drives model representations into distinct clusters.}
\label{fig:model_mds}
\end{figure}

To inspect the effect of face-stimulus-set optimization for each of the stimulus conditions, we compared model-layer representational geometries for randomly sampled stimuli and for controversial stimuli (\figurename~\ref{fig:model_mds}). Multidimensional scaling of layer-to-layer similarities shows that the face-stimulus sets optimized to be controversial among the models disentangle the model-predicted representational geometries. Each dot in the figure corresponds to a model layer, positioned so that distances between points reflect differences in the layers’ predicted vectors of representational dissimilarities. For random samples of faces (top row), different models made entangled predictions, with layers of different models predicting similar vectors of dissimilarities. For the controversial face sets (bottom row), each model layer predicts a representational geometry clearly distinct from all layers of all other models. Confusable model-predicted geometries occur only among different layers of the same model and mostly among layers of similar depth.

\begin{figure}[H]
\centering
\makebox[\linewidth][c]{\includegraphics[]{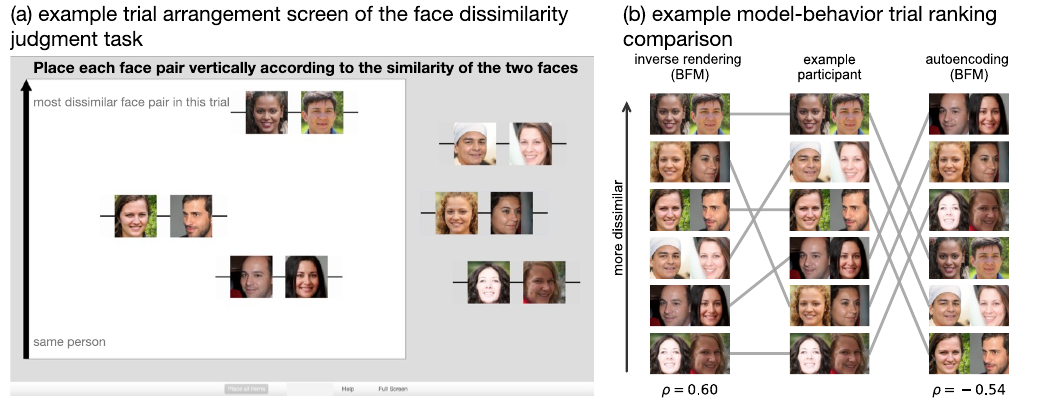}}
\caption{\textbf{Behavioral task and model--behavior comparison.}
\textbf{(a)} Example arrangement screen from the dissimilarity task. On each trial, participants dragged each face pair from the gray staging area into the white arena and positioned pairs vertically to indicate perceived dissimilarity, from ``same person'' (bottom) to ``most dissimilar face pair in this trial'' (top). 
\textbf{(b)} Illustrative model--behavior comparison. Face-pair rankings of the same trial from two example models (inverse rendering and autoencoding, both trained on BFM) are shown alongside an example participant's ranking. Gray lines connect the same face pair between model and participant rankings, illustrating how each model's predicted ordering corresponds to the participant's ordering. Spearman's rank correlation ($\rho$) quantifies the correspondence between a model's predictions and human judgments. The $\rho$ values shown are computed between the respective model and the participant's ranking for this trial. Our analysis was conducted by correlating model predictions with individual participant rankings on each trial before averaging across trials and participants.
}
\label{fig:trial_illustration}
\end{figure}

In all three experiments, we collected human dissimilarity judgments online (using Meadows and Prolific; Methods~\ref{sec:behavioral_task}). Participants arranged face pairs along a vertical dissimilarity axis by mouse drag-and-drop operations (Methods~\ref{sec:behavioral_task}). On each trial, participants arranged one set of six face pairs according to their perceived dissimilarity (\figurename~\ref{fig:trial_illustration}a). We quantified model performance at the trial level using Spearman's rank correlation $\rho$ between a model's predicted ordering of the six face-pair dissimilarities and the participant's ranking (\figurename~\ref{fig:trial_illustration}b). Evaluating models by their ordinal predictions rather than continuous representational dissimilarities intentionally discards information in order to make model performance insensitive to order-preserving transformations of the representational dissimilarities, while allowing direct model-human comparisons without additional fitted parameters. Since the best-predicting layer of each model was not known a priori, we treated layer identity as a nuisance parameter and used leave-one-subject-out cross-validation to select the best-predicting layer for each model (Methods~\ref{sec:statistical_inference_within}).

\subsection{Experiment 1: Frontal BFM stimuli}
In this experiment, we evaluated the models on frontal BFM faces. When testing on faces randomly sampled from the BFM latent space, all six models showed similar performance in predicting human dissimilarity judgments. Pairwise model comparisons were performed using a linear mixed-effects analysis of Fisher-$z$-transformed trial-wise correlations, followed by FDR correction; no model pair differed significantly (all $p_{\mathrm{corrected}}>.05$, \figurename~\ref{fig:frontal_bfm}, top row; see Supplementary Table~\ref{tab:model_comparison_cond} for full statistics; see Methods~\ref{sec:statistical_inference_within} for details of the model comparison and FDR correction). This outcome replicates the pattern reported in previous work using controlled synthetic face stimuli \cite{jozwik_face_2022}.

Controversial stimuli, designed to maximize disagreement between models, revealed more distinctions in model performance (\figurename~\ref{fig:frontal_bfm}, bottom row). In particular, two of the models trained on synthetic BFM faces, one for identification (faceID-BFM) and one for autoencoding (autoenc-BFM), were significantly outperformed by the other four candidate models. Models trained on natural images from VGGFace2 (faceID-VGGFace2 and autoenc-VGGFace2) consistently outperformed their synthetic-face-trained counterparts, for both the identification objective ($\Delta z=0.09, t=3.32, p_{\mathrm{corrected}}<.01$) and the autoencoding objective ($\Delta z=0.22, t=8.75, p_{\mathrm{corrected}} < .001$). These results suggest that exposure to real-world face-image statistics leads to more human-aligned representations. This better alignment is evident even when models were tested on synthetic BFM faces.

Furthermore, the inverse-rendering network (invRend-BFM), which is trained to invert the BFM generative process, also significantly surpassed the other BFM-trained networks (vs. faceID-BFM: $\Delta z=0.13, t=5.14, p_{\mathrm{corrected}} < .001$; vs. autoenc-BFM: $\Delta z=0.25, t=9.98, p_{\mathrm{corrected}} < .001$). This finding indicates that, within a synthetic training regime, learning to infer the underlying generative factors of stimuli yields representations that better predict face-dissimilarity judgments. Additionally, the objCat-ImageNet network outperformed faceID-BFM ($\Delta z=0.10, t=4.01, p_{\mathrm{corrected}} < .001$) and autoenc-BFM ($\Delta z=0.22, t=8.86, p_{\mathrm{corrected}} < .001$), indicating that broader object-level supervision can generalize to faces and yield human-aligned representations. Across all models, performance remained below the behavioral noise ceiling. Controversial sampling further widened the gap between the best model layers and the behavioral noise ceiling compared to random sampling. This indicates that none of the models fully explains the human judgments.

\begin{figure}[H]
\centering
\makebox[\linewidth][c]{\includegraphics[]{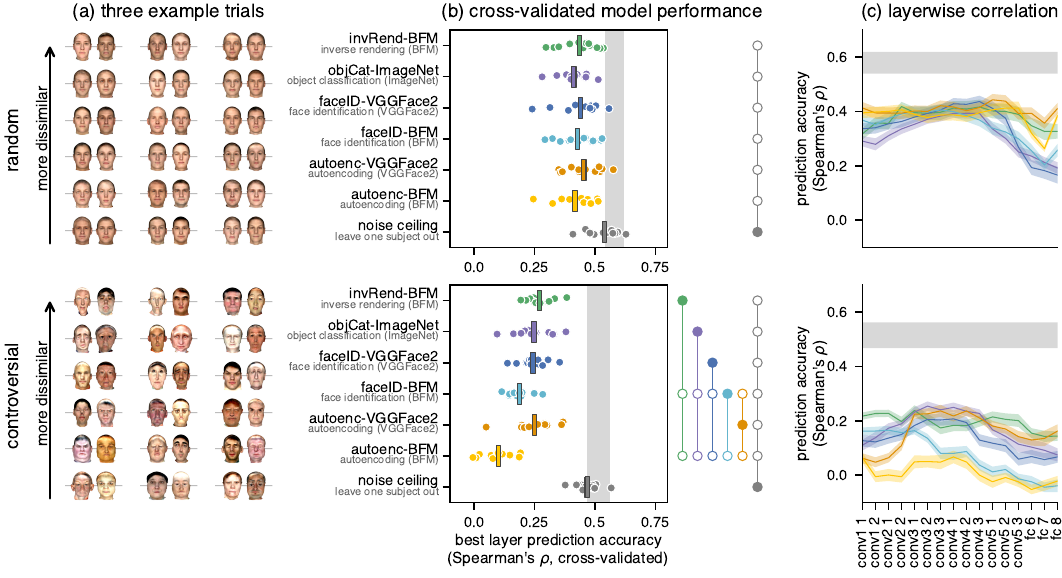}}
\caption{\textbf{Model performance on frontal BFM stimuli.} Candidate models were evaluated on their ability to predict human face-dissimilarity judgments for stimuli generated with random sampling (\textbf{top row}) or controversial stimulus synthesis (\textbf{bottom row}). 
\textbf{(a)} Three example trials (six face pairs per trial). The vertical arrangement of face pairs reflects the average dissimilarity rankings across participants. 
\textbf{(b)} Cross-validated model performance. Left: Each dot shows the average Spearman's $\rho$ correlation for one of the 12 independent stimulus sets (seeds). Within each set, the prediction accuracy of each model for each participant was computed using the model layer that best predicted the judgments of the other participants (leave-one-subject-out cross-validation). Bars show the mean across seeds. We report noise ceilings as the gray shaded region (lower and upper bounds) and include the leave-one-subject-out model to quantify the remaining gap between the candidate models and human behavior (Methods~\ref{sec:noise_ceiling}). Right: Statistical comparisons of model performance using metroplot (\url{https://github.com/brainsandmachines/metroplot}). A solid dot indicates that a model significantly outperformed the models connected by the open dots (linear mixed-effects model, $p<.05$, FDR corrected; see Methods, section~\ref{sec:statistical_inference_within}). 
\textbf{(c)} Layer-wise performance. Each colored line shows the performance of a single model across all its layers, and the shaded areas represent the standard error of the mean across different seeds (see Methods~\ref{sec:model_behavior_comparison}).}
\label{fig:frontal_bfm}
\end{figure}

\subsection{Experiment 2: Pose-varied BFM stimuli}
\begin{figure}[h!]
\centering
\makebox[\linewidth][c]{\includegraphics[]{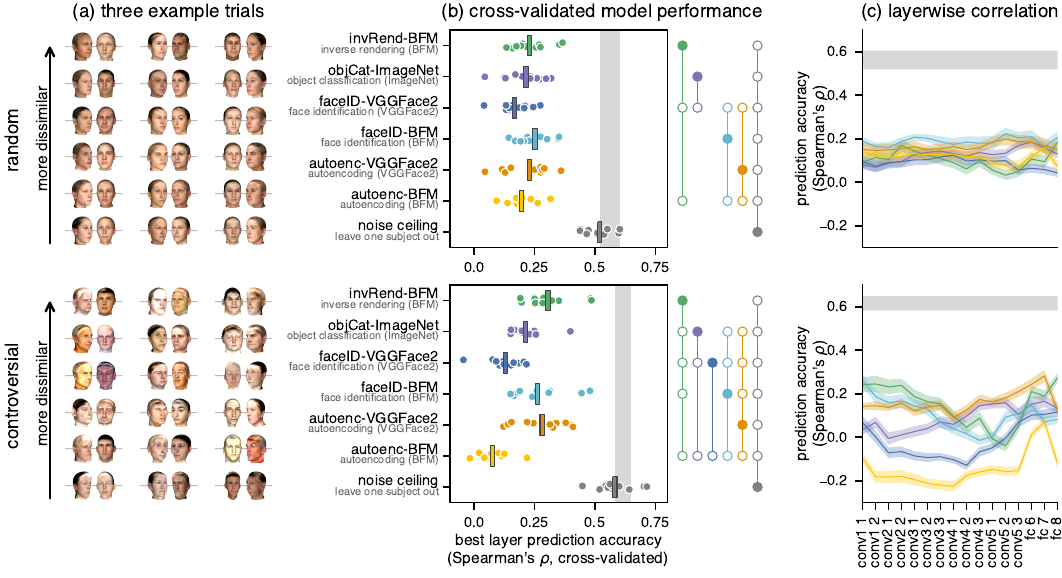}}
\caption{\textbf{Model performance on pose-varied BFM stimuli.} Panels and conventions match \figurename~\ref{fig:frontal_bfm}, but stimuli were BFM faces rendered with systematic head-pose variation. \textbf{(a)} Three example trials of face pairs that participants arranged. Note that within each trial, all six face pairs shared the same pose angles (one viewpoint for the left face and one for the right); viewpoint pairs vary across trials.
}\label{fig:pose_varied_bfm}
\end{figure}

Features used by our visual system to distinguish identities are expected to be sensitive to face shape and texture and relatively insensitive to nuisance variation, such as changes in viewpoint and lighting. When using only frontal faces, low-level features that are sensitive to both identity change and nuisance variation can be hard to disentangle from high-level face-identity features that are robust to nuisance variation \cite{carlin_adjudicating_2017, jozwik_face_2022}. We therefore introduced structured viewpoint variation using pose-varied BFM stimuli. Within each trial, all six face pairs were rendered from the same two viewpoints, with one viewpoint assigned to the left face and the other to the right face; this viewpoint pair varied across trials (\figurename~\ref{fig:pose_varied_bfm}a). The same pre-defined trial-specific viewpoint pairs were used for both randomly sampled and controversially optimized pose-varied BFM stimuli (Methods~\ref{sec:random_sampling}). This manipulation changes low-level image statistics while preserving face identity. 

With randomly sampled pose-varied BFM faces, model differences became more evident (\figurename~\ref{fig:pose_varied_bfm}b,c). Three models performed best: invRend-BFM, faceID-BFM, and autoenc-VGGFace2. Each significantly outperformed faceID-VGGFace2 and autoenc-BFM (all $p_{\mathrm{corrected}} < .05$, see Supplementary Table~\ref{tab:model_comparison_cond}). Additionally, objCat-ImageNet significantly outperformed faceID-VGGFace2 ($\Delta z=0.07, t=2.54, p_{\mathrm{corrected}}<.02$). As before, controversial stimuli further increased model discriminability. invRend-BFM and autoenc-VGGFace2 emerged as the two best-performing candidates and outperformed all other models (all $p_{\mathrm{corrected}} < .05$, see Supplementary Table~\ref{tab:model_comparison_cond}). invRend-BFM and autoenc-VGGFace2 were among the best-performing models in the frontal BFM experiment, indicating that their strong performance generalized across BFM stimulus sets with and without pose variation.

To summarize the results of pose-varied BFM stimuli, we conducted a model-dominance analysis across both random and controversial sampling conditions. Model $A$ was considered to strictly dominate model $B$ if $A$ performed at least as well as $B$ in both sampling strategies and strictly better in at least one. This analysis revealed the same ranking as observed in the controversial condition alone, reflecting consistent relative model performance across sampling methods. This consistency arose because any model that significantly outperformed another under random sampling continued to do so under controversial stimulus synthesis, with performance differences often becoming more pronounced. For instance, the performance gap between the BFM-trained models invRend-BFM and autoenc-BFM widened from the random condition ($\Delta z=0.07, t=2.60, p_{\mathrm{corrected}}<.02$) to the controversial condition ($\Delta z=0.31, t=11.70, p_{\mathrm{corrected}}<.001$). Overall, controversial synthesis increased inferential power without qualitatively reordering model performance within this stimulus family. Relative to Experiment 1, faceID-BFM moved from the lower-performing group in the frontal test condition to third place when tested with pose variation. This suggests that learning to identify faces of varying poses within the same image domain as the test stimuli (i.e., BFM faces) has advantages for handling pose variation. However, learning to identify BFM faces remained less effective than learning to infer generative factors (invRend-BFM) or learning a natural-image distribution (autoenc-VGGFace2). autoenc-VGGFace2 readily generalized to the more controlled setting of BFM faces.

\subsection{Experiment 3: StyleGAN3 stimuli}
Finally, we evaluated the models on photorealistic faces generated by StyleGAN3, which introduces rich diversity and realism at the expense of losing explicit control over factors such as pose and lighting, while avoiding familiarity confounds inherent to celebrity face datasets. With randomly sampled StyleGAN3 faces, faceID-VGGFace2 significantly outperformed all five other models (\figurename~\ref{fig:stylegan3_main_results}, all $p_{\mathrm{corrected}} < .001$, see Supplementary Table~\ref{tab:model_comparison_cond}). This result suggests that the diversity in the test stimuli was sufficient to separate our model candidates, even without controversial optimization. Controversial stimuli reinforced this pattern and further highlighted the advantage of discriminative training on real images. Both faceID-VGGFace2 and objCat-ImageNet significantly outperformed the four other models (all $p_{\mathrm{corrected}} < .001$, see Supplementary Table~\ref{tab:model_comparison_cond}). 

The model-dominance analysis across the two sampling methods showed that faceID-VGGFace2 remained the strongest overall performer, suggesting that face-specific identity supervision captured aspects of human face-similarity judgments that were not fully explained by object-level supervision. Yet, the strong performance of objCat-ImageNet indicates that a substantial component of representational geometry underlying face dissimilarity judgments is shared with features learned for general visual categorization.

In contrast to the pose-varied BFM results of Experiment 2, where inverse-generative models were favored, discriminative models trained on natural images performed the best on StyleGAN3 stimuli. This divergence points to an important interaction between the stimulus domain and the training realism of the models: conclusions about model performance drawn from results in a single face-stimulus domain do not necessarily generalize to other domains with different statistics and sources of variation.

\begin{figure}[H]
\centering
\makebox[\linewidth][c]{\includegraphics[]{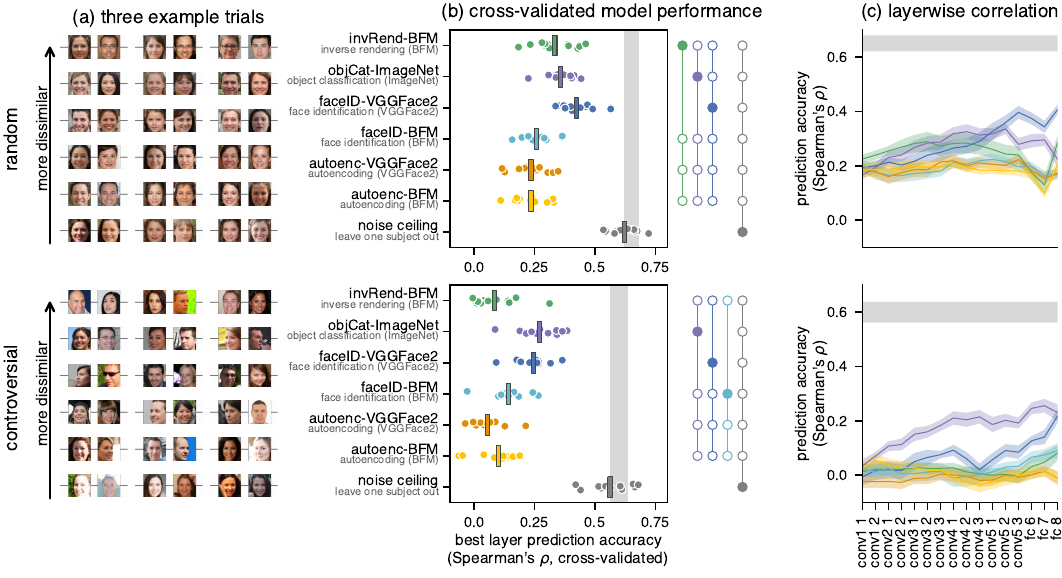}}
\caption{\textbf{Model performance on photorealistic StyleGAN3 stimuli.} Panels and conventions match \figurename~\ref{fig:frontal_bfm}, but stimuli were generated by StyleGAN3 \citep{karras_alias-free_2021}. Unlike BFM faces, StyleGAN3 stimuli have photorealistic appearance variation (including background and illumination), providing a complementary test of model generalization.
}\label{fig:stylegan3_main_results}
\end{figure}

\subsection{Summary of condition-specific results}
Across three distinct stimulus families, introducing systematic variability in pose or facial features consistently improved our ability to distinguish among competing computational models. Within each family, the relative ordering of models was largely preserved between random sampling and controversial stimulus synthesis, while controversial stimuli generally increased performance differences.

A central finding from these analyses is that no single model consistently outperformed all others in all conditions. Instead, the top-performing model depended on the test stimuli, showing that there is an interaction between stimulus realism and diversity on one hand and the model's training regime on the other. On BFM stimuli (both frontal and pose-varied), the inverse-rendering model and autoenc-VGGFace2 were consistently among the best models. This pattern suggests that learning to represent the latent generative factors---either explicitly (inverse rendering) or implicitly (autoencoding)---gives rise to representations that align well with human judgments in these synthetic contexts. In contrast, on photorealistic faces generated by StyleGAN3, the faceID-VGGFace2 model performed best. This result shows that identity supervision on real photos aligns well with human face-similarity judgments of naturalistic stimuli.

Across stimulus families and training regimes, the best-performing layer within VGG-16 varied substantially. For frontal BFM stimuli, mid-level convolutional layers predicted human judgments best. For pose-varied BFM stimuli, these same mid-level layers performed worse, especially for controversial stimuli: some models favored early layers while others favored late layers. For StyleGAN3 stimuli, late layers performed best in the two networks that achieved good overall performance. Taken together, these shifts in the winning layer across conditions are incompatible with the hypothesis that any single layer of the tested networks corresponds to the representational space underlying human similarity judgments. These results show that using a wide variety of stimuli (including photorealistic stimuli with naturalistic levels of nuisance variation) is essential for the assessment of computational models of human face perception. This pattern further suggests that human perceptual judgments may reflect a readout mechanism that draws from both lower and higher levels of the representational hierarchy.

\subsection{Condition-pooled mixed-effects analysis}

\begin{figure}[b]
\centering
\makebox[\linewidth][c]{\includegraphics[]{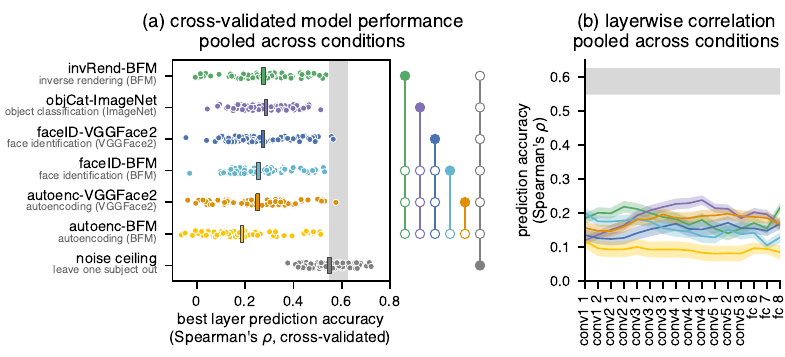}}
\caption{\textbf{Model performance pooled across conditions.} \textbf{(a)} Cross-validated model performance pooled across all six conditions (three stimulus families $\times$ two sampling strategies). Model performance was quantified as Spearman's rank correlation between dissimilarity judgments and model predictions using the most predictive layer. To avoid selection bias, performance was estimated using leave-one-subject-out cross-validation among participants who viewed the same stimulus set. Each dot represents the average performance for a single random seed within one of the six experimental conditions ($N=72$ dots per model). The metroplot (right) visualizes statistical significance: a solid colored dot indicates that a model significantly outperformed the models indicated by the connected open dots (linear mixed-effects model, $p<.05$, FDR corrected; see Methods~\ref{sec:statistical_inference_pooled}). The complete set of pairwise model-comparison statistics for this pooled analysis is reported in Supplementary Table~\ref{tab:model_comparison_pooled}.
\textbf{(b)} Layer-wise performance pooled across conditions. We computed average performance of every layer in each model, pooled across all participants, trials, seeds, and conditions. Shaded regions represent the standard error of the mean (SEM), calculated across the 72 unique seed-by-condition combinations.
}\label{fig:pooled_performance}
\end{figure}

While condition-wise analyses reveal specific modeling strengths, it is useful to determine which models have more robust representations across a variety of face stimuli and experimental designs. We therefore performed a pooled mixed-effects analysis that combined model--behavior correlations across all six conditions (three stimulus families $\times$ two sampling strategies) and compared models using a linear mixed model that accounts for the hierarchical structure of the data (Methods~\ref{sec:statistical_inference_pooled}). This analysis revealed three models that significantly outperformed the remaining candidates: invRend-BFM, faceID-VGGFace2, and objCat-ImageNet (\figurename~\ref{fig:pooled_performance}). These three winning models were statistically indistinguishable from each other. Thus, the most robust performance across conditions was achieved by models optimized either for inverse-generative inference or for discriminative learning on natural images.

\section{Discussion}\label{sec12}

Our results suggest that human face perception may be optimized both to infer latent face variables by inverting the image generation process and to compute discriminative features that discard naturalistic nuisance variation. 

Methodologically, our analyses of six models across a wide range of conditions demonstrate the importance of naturalistic, varied, and optimized stimuli for revealing shortcomings of current models. This thorough experimental approach reveals that all models have a large gap to the noise ceiling, leaving substantial variance unexplained. Our study also demonstrates the need to consider a range of models in order to avoid premature identification of a winning model, and to design experiments that can adjudicate among competing hypotheses \citep{golan_controversial_2020}. Only through using a rich variety of stimuli and models can we appreciate the failures of our current theories and their implementations in models, and eventually discover the computational mechanisms of human face perception.

Concretely, our experiments demonstrate that introducing task-relevant naturalistic stimulus variability and using synthetic controversial stimuli \citep{golan_controversial_2020, golan_testing_2023} improved the inferential power of model comparison for human-face dissimilarity judgments. Pose variation and photorealistic diversity better differentiated model representational geometries, even under random sampling. Controversial stimulus synthesis further magnified differences in model performance while largely preserving their relative rank order within each stimulus family. Although none of the models reached the behavioral noise ceiling, our findings reveal key factors that shape the alignment of model representations with human perception, allowing us to address the guiding questions posed in the Introduction.

\textbf{What computational objectives best explain human face-dissimilarity judgments?} Pooled across conditions, three models outperformed the remaining candidates: inverse rendering \citep{yildirim_efficient_2020, ryali_leveraging_2020}, face identification trained on natural photographs \citep{dobs_behavioral_2023, jiahui_modeling_2023, tian_face_2022, jacob_qualitative_2021, yovel_deep_2023, abudarham_critical_2019, BLAUCH2021104341, rosemblaum_concurrent_2025}, and domain-general object categorization \citep{grossman_convergent_2019, vinken2023, chang_explaining_2021}. These models share critical features: they prioritize identity-relevant high-level representations robust to nuisance variation. The inverse rendering objective infers the latent generative factors of face shape and texture independently of view and lighting. Discriminative classification on natural images also requires the model to learn transformation-invariant representations that enable generalization across views. Robust generalization across a variety of face stimulus sets appears to require either explicit generative modeling or supervision on natural image statistics. While autoencoding models---which implicitly learn generative factors---performed well in specific synthetic conditions, they failed to generalize overall. One possibility is that their pixel-reconstruction objective forces them to encode low-level image statistics, which are less predictive of perceived similarity under complex visual variation \citep{daube_grounding_2021}. In contrast, the three winning models all prioritize high-level structures or semantic content over pixel fidelity.

\textbf{What visual diet for model training yields human-like face representations?} Ecological realism in the training diet is crucial for generalization: For a given training objective, models trained on natural photographs typically outperformed synthetic-trained counterparts, even when evaluated on synthetic face stimuli. This asymmetry suggests that a purely synthetic diet may omit visual dimensions that human perception uses reliably (e.g., realistic textures, complex illumination, and external features such as hair \citep{abudarham_critical_2019}). Furthermore, generalization is not uniform across test domains: model rankings differed between BFM and StyleGAN3 stimuli, indicating that conclusions about a single best model are fragile when stimulus realism, diversity, and nuisance structure vary. These findings motivate model evaluation across multiple stimulus domains that jointly span a range of tradeoffs between experimental control and ecological realism. In particular, 3D morphable models like BFM remain essential for controlled manipulations of pose and illumination. However, photorealistic generators are also essential because they expand diversity and more closely approximate real-world statistics. 

\textbf{Which models best explain human dissimilarity judgments under face-pose variation?} Pose variation is an informative manipulation because head pose is a nuisance transformation: it can strongly alter low-level image statistics while leaving identity unchanged. At the same time, the degree to which humans achieve viewpoint-invariant face recognition, especially for unfamiliar faces, remains debated \citep{troje_face_1996, bruce_verification_1999, jenkins_variability_2011, young_burton_2017, grossman_convergent_2019, parde_twin_2023, zhu_view_2025}. In our pose-varied BFM condition, models that explicitly or implicitly factorize generative factors aligned most closely with human judgments: the inverse rendering model and the autoencoding model trained on natural photographs were consistently among the best-performing models. Inverse rendering explicitly separates pose and lighting from identity-related latents such as shape and texture, whereas an autoencoder on diverse natural images implicitly encodes systematic variation of appearance (including pose) in order to be able to reconstruct the input image. Our results suggest that human perception deemphasizes viewpoint-related variation in face-similarity judgments. Consistent with this, a previous study found that pose and camera parameters account for little variance in triplet odd-one-out judgments of both static and dynamic faces \citep{Hofmann_2025}. By contrast, discriminative face-identification objectives did not provide the strongest account of human judgments under pose variation. Face identification models rely heavily on viewpoint-dependent cues rather than robust 3D representations \citep{alcorn_strike_2019}. As a result, despite being trained to map to identities, these models are not invariant to pose changes \citep{hill_deep_2019}.

\textbf{Limitations and future directions.} Two deliberate design choices limit the scope of the present conclusions. First, we compared training objectives and visual diets within a single feedforward CNN architecture, VGG-16 \citep{simonyan_vgg16_2015}. This design isolated the effects of training objective and visual diet, but leaves open how well the findings generalize across architectures. We focused on CNNs because their hierarchically organized, spatially local computations, together with receptive fields that expand across depth, provide a more neurally grounded approximation to primate ventral-stream processing \citep[e.g.][]{yamins_performance-optimized_2014, khaligh-razavi_deep_2014, guclu_deep_2015, kubilius_brain-like_2019, lindsay_convolutional_2021} than current transformer-based alternatives. Nevertheless, fixing architecture limits the generality of the conclusions that can be drawn. A particularly important extension is to recurrent models, which better capture the temporal dynamics of ventral-stream object recognition \citep[e.g.][]{spoerer_recurrent_2017, spoerer_recurrent_2020, kietzmann_recurrence_2019, kar_evidence_2019, rajaei_beyond_2019} and may reveal mechanisms---such as iterative refinement of shape and texture estimates, and dynamic switching from face detection to face identification \citep{shi_rapid_2026}---that purely feedforward models cannot capture. A related limitation is that each computational objective was instantiated by a single model, training objective, and training domain, and that implementation choices were not fully factorial. Thus, some condition-specific differences may reflect training--test domain similarity or implementation details in addition to the intended objective; indeed, invRend-BFM performed particularly well on BFM stimuli, whereas faceID-VGGFace2 performed best on StyleGAN3 stimuli. Second, we trained the models solely on perceptual tasks and did not fit them, linearly or nonlinearly, to human judgments. Our aim was to test whether training objective and visual diet alone give rise to representations that generalize out-of-distribution to human behavior. Prior work shows that training and especially fitting can substantially improve DNN-human alignment \citep[e.g.][]{storrs_2021_diverse, muttenthaler2023human, avitan2025modelbehavior}, but may obscure the contribution of the original optimization objective. The gap between even our best-performing models and the human noise ceiling therefore constitutes a meaningful target for theory, but closing it will also depend on implementation choices---including architecture and the mapping from model representations to behavior. Looking forward, important next steps are to broaden both the architectural and objective spaces---particularly by incorporating recurrent computations and richer, potentially combined, training objectives \citep{moscovitch_what_1997, plaut_behrmann_2011, otoole_face_2021, kar_deep_2023, shoham_visual_semantic_2024, grosbard_yovel_2025, lee_efficient_2026}---and to test models jointly against behavioral judgments and neural measurements from face-selective cortex \citep{kanwisher_fusiform_1997, freiwald_tsao_2010} to ask whether the objectives that best explain perceptual judgments also best explain cortical representations.

\section{Methods}

\subsection{Model training}
\label{sec:model_training}
All candidate models were built upon the VGG-16 architecture \citep{simonyan_vgg16_2015} (torchvision implementation) without batch normalization, operating on $128\times 128$-pixel RGB inputs. Model weights were initialized using Kaiming normal initialization \cite{he_init_2015}. For each training dataset, input images were standardized using per-channel statistics computed from that dataset, and the same normalization was applied during stimulus optimization and model--behavior analyses. Models were trained using PyTorch Lightning \citep{Falcon_PyTorch_Lightning_2019}, with dropout, where present, enabled only during training.

\subsubsection{faceID-BFM}

The faceID-BFM model was trained as a supervised face-identification model on a synthetic BFM-identity dataset. We first sampled 8,631 synthetic identities (matching the number of training identities in VGGFace2) by drawing one unique set of BFM shape and texture coefficients for each identity. For each identity, we rendered 363 images while holding the identity-defining shape and texture coefficients fixed and varying nuisance variables including expression, pose, light direction, and light intensity. This procedure yielded approximately 3.13 million images, close to the number of images in the VGGFace2 training set \citep{cao_vggface2_2018}. 

The model was trained to classify each image according to its synthetic identity label using a cross-entropy loss. The output layer was replaced with an identity classification readout. During training, images were randomly cropped, normalized using per-channel statistics, and augmented using random grayscale conversion ($p=0.1$), photometric perturbations, noise, occlusion-like dropout, geometric distortions, blur-related transformations, and horizontal flips \citep{buslaev_albumentations_2020}.

\subsubsection{faceID-VGGFace2}

The faceID-VGGFace2 model was trained as a supervised face-identification model on real face photographs from VGGFace2 \citep{cao_vggface2_2018}. The model was optimized using cross-entropy loss to predict the person identity associated with each training-set image.

Because VGGFace2 contains different numbers of images per identity, training examples were sampled using an identity-balancing weighted sampler. Training images were cropped using the VGGFace2 face bounding boxes, resized, and randomly cropped during training. Random grayscale conversion with probability of 0.2 was applied during training.

\subsubsection{invRend-BFM}

The invRend-BFM model was trained to invert the BFM rendering process. We generated approximately 3.30 million synthetic face images by independently sampling the BFM generative variables: shape, texture, expression, pose, light direction, and light intensity. The supervised target for each image was the concatenated vector of sampled variables used to render it. Unlike the BFM-identity dataset used for faceID-BFM, this dataset did not contain multiple nuisance-varied images per synthetic identity, thereby reducing the opportunity for shortcut learning based on memorizing identity-specific latent codes.

For each training image $\mathbf{x}_i$ and its generative-parameter vector $\mathbf{y}_i$, the network predicted a 508-dimensional generative-parameter vector $\hat{\mathbf{y}}_i = f_{\theta}(\mathbf{x}_i)$, partitioned into six BFM parameter groups: shape coefficients ($d=199$), texture coefficients ($d=199$), expression coefficients ($d=100$), pose quaternion ($d=4$), light intensity ($d=3$), and light direction ($d=3$). To prevent the high-dimensional shape, texture, and expression blocks from dominating the low-dimensional nuisance variables, we used a group-normalized mean-squared-error loss:
\begin{equation}
\mathcal{L}_{\mathrm{invRend}} =
\frac{1}{N}
\sum_{i=1}^{N}
\sum_{g \in \mathcal{G}}
\frac{1}{d_g}
\sum_{k=1}^{d_g}
\frac{
\left(\hat{y}_{i,g,k} - y_{i,g,k}\right)^2
}{
\sigma_{g,k}^2
},
\end{equation}
where $N$ is the number of images in a minibatch, $\mathcal{G}$ denotes the six parameter groups, $d_g$ is the dimensionality of group $g$, and $\sigma_{g,k}^2$ is the normalizing variance for dimension $k$ of group $g$. For shape, texture, and expression coefficients, $\sigma_{g,k}^2$ was set to 1, reflecting the unit-variance Gaussian prior for BFM latents; for pose, light intensity, and light direction, $\sigma_{g,k}^2$ was set to the empirical variance of the corresponding target variable in the training set.

During training, rendered heads were cropped from the background, resized, randomly cropped, and converted to grayscale with probability 0.2.

\subsubsection{autoenc-BFM and autoenc-VGGFace2}
The autoenc-BFM and autoenc-VGGFace2 models were trained as variational autoencoders \citep{kingma_auto-encoding_2014} to reconstruct their input images while regularizing the latent distribution toward a unit Gaussian prior. This objective provided an unsupervised generative-learning counterpart to the supervised face-identification and inverse-rendering objectives.

We used the SigmaVAE formulation \citep{rybkin_sigmavae_2020}, in which the standard deviation of the Gaussian residual term is learned rather than predetermined as a hyperparameter. The final VGG-16 task-specific readout was replaced by a linear readout that produced two 500-element vectors: a mean vector and a log-variance vector, defining a 500-dimensional Gaussian latent distribution with diagonal covariance. During training, a learned decoder mapped samples from this latent distribution back to $128\times128$ RGB images by projecting them to a $256\times8\times8$ feature map, applying four convolutional upsampling stages, and using a final convolution followed by a sigmoid output. The reconstruction term was defined as the Gaussian negative log-likelihood of the input image under a Gaussian centered at the decoded image, with a single learned scalar parameter $\sigma$ specifying the residual standard deviation, shared across pixels, channels, and images. The full loss was the sum of this reconstruction term and a $\beta$-weighted KL divergence term regularizing the approximate posterior toward a unit Gaussian prior, with $\beta=1$:
\begin{equation}
\mathcal{L}_{\mathrm{VAE}} =
-\log p_{\theta,\sigma}(\mathbf{x} \mid \mathbf{z}) +
\beta D_{\mathrm{KL}}\left(q_{\phi}(\mathbf{z} \mid \mathbf{x}) \,\|\, p(\mathbf{z})\right).
\end{equation}

After training, the decoder and log-variance projection were discarded. For stimulus optimization and model--behavior analyses, we used the trained encoder only, with the deterministic posterior mean serving as the autoencoder's final latent representation. Intermediate encoder layers were analyzed analogously to the corresponding layers in the other VGG-16-based models.

The autoenc-BFM model was trained on the same synthetic BFM image set used for training the invRend-BFM model, excluding the associated generative-parameter labels. The autoenc-VGGFace2 model was trained on the VGGFace2 training set, excluding the identity labels. For BFM images, preprocessing and augmentation matched the synthetic BFM image pipeline used for invRend-BFM. For VGGFace2 images, preprocessing and augmentation matched the faceID-VGGFace2 pipeline.

\subsubsection{objCat-ImageNet}
The objCat-ImageNet model was trained as an object-categorization model on the ImageNet-1K/ILSVRC2012 training set \citep{deng_imagenet_2009}. The VGG-16 classifier output layer was replaced with a 1,000-way ImageNet category readout, and the model was optimized using cross-entropy loss to predict object-category labels. Training images were augmented using random resized crops, horizontal flips, and grayscale conversion with probability 0.2.

\subsubsection{Model instances}
For each candidate model, we trained three independent instances from different random weight initializations. These instances had the same architecture, training objective, and training dataset, but differed in their random initialization. One instance was used for generating ground-truth responses during stimulus optimization (the data-generating instance), a second was used as a reference instance when computing within-model similarity and comparing model representations to human behavior, and a third was held out for model recovery experiments (Methods~\ref{sec:model_recovery}). This separation allowed us to distinguish model-level representational predictions from idiosyncrasies of a single trained network instance. Complete training hyperparameters and validation metrics are provided in Supplementary Table~\ref{tab:model_training_summary}.

\subsection{Stimulus generation}
\subsubsection{Face generators}
We used two face generators to construct the experimental stimulus sets: Basel Face Model (BFM 2019, a 3D morphable model; \cite{bfm17, bfm09, blanz_morphable_1999}) and StyleGAN3 \cite{karras_alias-free_2021}. BFM 2019 parameterizes face identity in separate shape and texture latent spaces, and provides precise control over nuisance variables such as face pose, lighting direction, and lighting intensity. To make BFM differentiable for stimulus optimization, we used PyTorch3D \citep{ravi2020pytorch3d} to implement the rendering process.

StyleGAN3, in contrast, generates highly naturalistic faces and avoids the familiarity confounds in celebrity face datasets, but does not provide explicit control over pose and background. To reduce GPU memory demands during stimulus optimization, we retrained StyleGAN3 to generate faces at 128 $\times$ 128 pixels, matching the resolution used by the six representational models.

\subsubsection{Generating randomly sampled face sets}
\label{sec:random_sampling}
For each experimental seed and stimulus family, we randomly sampled 144 face images, forming 72 pairs arranged into 12 trials with six pairs per trial. The family-specific stimulus-set construction procedures were as follows.

\paragraph{Frontal BFM stimuli} Each face identity was specified by a shape latent vector, $\boldsymbol{\alpha}_s \in \mathbb{R}^{199}$, and a texture latent vector, $\boldsymbol{\alpha}_t \in \mathbb{R}^{199}$, with both vectors drawn from $\mathcal{N}(\boldsymbol{0},\mathbf{I})$. We rendered these faces in a frontal view under identical ambient lighting. The faces were then randomly paired, and the resulting pairs were randomly grouped into trials.

\begin{figure}[h]
\centering
\makebox[\linewidth][c]{\includegraphics[scale=0.75]{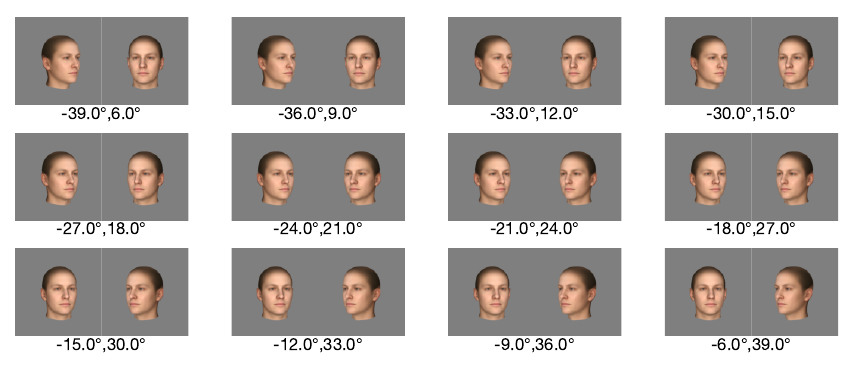}}
\caption{\textbf{Predefined face-pose angles, with one angle pair assigned to each trial}. We varied face pose only along the yaw axis. Within each trial, all six face pairs shared the same pose configuration, with one yaw angle assigned to the left face and another to the right face. Because each stimulus set comprised 12 unique trials, we defined 12 pose configurations in total.}
\label{fig:trial_angles}
\end{figure}

\paragraph{Pose-varied BFM stimuli} 
Sampling and pairing were identical to the frontal condition, but head pose was systematically varied. We defined 12 trial-level yaw configurations, each specifying a 45$^\circ$ separation between the left and right faces: $(\alpha,\ \alpha+45^\circ)$ with $\alpha \in \{-39^\circ,-36^\circ,\ldots,-9^\circ,-6^\circ\}$. These configurations ranged from $(-39^\circ,6^\circ)$ to $(-6^\circ,39^\circ)$, with the two faces oriented toward one another (\figurename~\ref{fig:trial_angles}). A single yaw configuration was assigned to each trial, so all six face pairs within a trial shared the same pose configuration.

\paragraph{StyleGAN3 stimuli}
We sampled latent vectors from the StyleGAN3 $Z$ space, drawing from the standard normal distribution, $\mathcal{N}(\boldsymbol{0},\mathbf{I})$, and used the generator to synthesize corresponding face images. As in the frontal BFM stimulus family, the synthesized images were randomly paired, and the resulting pairs were randomly grouped into trials.

\subsubsection{Generating controversial face sets}
\label{sec:controversial_face_synthesis}
\paragraph{Model discrimination objective}
To generate controversial face sets that efficiently discriminate among our candidate models, we optimized the stimulus set using a Bayesian optimal experimental design approach \citep{Chaloner_1995_Bayesian, myung_optimal_2009, cavagnaro_myung_adaptive_2010}. Rather than optimizing protocol variables such as presentation order or timing, we optimized the stimulus set $\xi$ with respect to a utility function that maximizes the differences in candidate models' representational geometries \citep{golan22}. See also recent work optimizing stimuli to maximize between-model differences in \emph{local} representational geometry around individual images \citep{feather2025discriminating}. Here, we focus on \emph{global} representational geometry: the set of pairwise representational dissimilarities among the presented stimuli.

We first present a simple objective that quantifies how well an experimental design $\xi$ discriminates among $M$ candidate models (Eqs.~\ref{eq:local_utility}--\ref{eq:global_utility}). We then generalize this objective to handle neural network models with multiple candidate layers (Eqs.~\ref{eq:multi_layer_local_objective}--\ref{eq:multi_layer_global_objective}). Both objectives can be paired with any model performance estimator $\psi(\cdot, \cdot \mid \xi)$ used to compare model predictions with empirical data. We describe the specific model performance estimator we used (Eq.~\ref{eq:model_performance_estimator_pearson}) below.

\subparagraph{Discriminating between $M$ models} Given $M$ candidate models, we treat each model in turn as the data-generating model and measure how well its predicted responses for stimulus set $\xi$ would be distinguishable from the predictions of every alternative model. For a given data-generating model $m$ we define the \emph{local utility} $u(\xi, m)$ as
\begin{equation}
u(\xi,m) = f\!\Big[
\underbrace{\psi\!\big(\tilde{\mathbf{y}}^{m},\,\hat{\mathbf{y}}^{m}\mid \xi\big)}_{\mathclap{\substack{\text{how well model $m$ predicts}\\\text{data generated by a held-out}\\\text{instance of itself}}}}
\;\;-\;\;
\underbrace{\max_{m' \neq m}\psi\!\big(\tilde{\mathbf{y}}^{m},\,\hat{\mathbf{y}}^{m'}\mid \xi\big)}_{\mathclap{\substack{\text{how well the best} \\ \text{alternative model predicts} \\ \text{data generated by model } m}}}
\Big],
\label{eq:local_utility}
\end{equation}
where $\tilde{\mathbf{y}}^m$ is the vector of dissimilarity predictions produced by a data-generating instance of model $m$ for stimulus set $\xi$, and $\hat{\mathbf{y}}^m$ is the corresponding prediction made by a reference instance of the same model (identical architecture, training objective, and training data, but initialized with a different random seed). Similarly, $\hat{\mathbf{y}}^{m'}$ is the prediction made by a reference instance of an alternative model $m'$. $\psi(\cdot, \cdot \mid \xi)$ is a differentiable model-performance estimator (specified below), and $f$ is a monotonically increasing function. We take $f(x) = -e^{-10x}$, which emphasizes incorrect model recoveries (large negative differences) and de-emphasizes cases in which large model discrimination has already been achieved (large positive differences). In a prior simulation study this choice yielded high recovery accuracy of the true data-generating model, relative to alternatives such as the identity function \citep{golan22}.

The \emph{global utility} is the expectation of the local utility under our prior over candidate data-generating models,
\begin{equation}
U(\xi) \;=\; \sum_{m=1}^{M} p(m)\, u(\xi, m),
\label{eq:global_utility}
\end{equation}
where $p(m)$ encodes our prior belief that model $m$ is the true data-generating process (uniform prior to any empirical data). $U(\xi)$ captures the expected advantage of a data-generating model over the strongest alternative candidate model.

If the first term is omitted from the local utility (Eq. \ref{eq:local_utility}), the global utility (Eq. \ref{eq:global_utility}) reduces to a simpler criterion: selecting stimuli whose candidate-model predictions are as distinct as possible. Stimulus-selection objectives that decorrelate representational predictions \citep{jozwik_disentangling_2022,Hosseini2024universality} fall into this class. Including the first term accounts for the expected reliability of the measurements.

\subparagraph{Accounting for multiple neural network layers} A common approach is to evaluate each network by its most human-aligned layer \citep[e.g.,][]{schrimpf_integrative_2020, storrs_2021_diverse, conwell_large-scale_2024}. During stimulus optimization, before any empirical data are available, we could treat each layer as a separate candidate in Eq.~\ref{eq:global_utility}; doing so, however, would devote much of the stimulus design's power to discriminating among consecutive layers of the same network, which typically yield closely related representations. Because our goal is to adjudicate among entire networks rather than among their layers, we instead treat layer identity as an unknown nuisance parameter, yielding the following local utility:
\begin{equation}
u(\xi, m,l) = f\Big[ \max_{l'} \psi\big(\tilde{\mathbf{y}}^{m,l}, \hat{\mathbf{y}}^{m,l'} \mid \xi \big) - \max_{m' \neq m} \max_{l'} \psi\big(\tilde{\mathbf{y}}^{m,l}, \hat{\mathbf{y}}^{m',l'}\mid \xi \big) \Big],
\label{eq:multi_layer_local_objective}
\end{equation}
where $\tilde{\mathbf{y}}^{m,l}$ is the vector of dissimilarity predictions produced by layer $l$ of the data-generating instance of model $m$ and $\hat{\mathbf{y}}^{m,l'}$ is the vector of dissimilarity predictions produced by layer $l'$ of the reference instance of the same model. Similarly, $\hat{\mathbf{y}}^{m',l'}$ is the dissimilarity-prediction vector produced by layer $l'$ of the reference instance of an alternative model $m'$.

This local utility is marginalized over data-generating models and over candidate layers within each data-generating model to yield a global utility measure:
\begin{equation}
U(\xi) = \sum_m p(m) \sum_{l=1}^{L_m} p(l \mid m)\, u(\xi, m,l),
\label{eq:multi_layer_global_objective}
\end{equation}
where $p(l \mid m)$ is the prior probability that layer $l$ is the data-generating representation given model $m$ (uniform across the $L_m$ layers of model $m$). As in Eq.~\ref{eq:global_utility}, we compare the data-generating model against the best-performing alternative model and additionally take the best-matching layer within each model. The reference instances, marked with hats, are reused in the subsequent data analysis to evaluate model performance on the empirical judgments.

\subparagraph{Model performance estimator}

The utility function depends on a model-performance estimator $\psi(\cdot, \cdot \mid \xi)$ that measures how well one dissimilarity vector predicts another over stimulus set $\xi$. The estimator used in stimulus optimization should match the measure planned for evaluating model--behavior alignment on the empirical data as closely as possible, while also being differentiable with respect to $\xi$ to support gradient-based stimulus optimization. 

Because the behavioral task asked participants to sort face pairs separately within each trial, we used a corresponding trial-wise model-performance estimator. Specifically, in the analysis of the behavioral experiment, we quantified the performance of candidate model $m'$ at layer $l'$ by first computing, within each trial, Spearman's rank correlation between the participant's dissimilarity judgments and the model-predicted dissimilarities for the face pairs in that trial, and then averaging these correlation coefficients across the $T=12$ trials:
\begin{equation}
\psi_\rho\big(\mathbf{y}, \hat{\mathbf{y}}^{m',l'} \mid \xi\big)
= \frac{1}{T} \sum_{t=1}^{T} \rho\big(\mathbf{y}_{t}, \hat{\mathbf{y}}^{m',l'}_{t}),
\label{eq:model_performance_estimator_rho}
\end{equation}
where $\mathbf{y}_{t}$ is the vector of participant dissimilarity judgments for trial $t$ in stimulus set $\xi$, and $\hat{\mathbf{y}}^{m',l'}_{t}$ is the corresponding vector of predicted dissimilarities produced by layer $l'$ of the reference instance of model $m'$ for that trial. Model-predicted dissimilarities were computed as the squared Euclidean distances between the flattened layer-activation vectors (post-ReLU, where applicable) elicited by the two faces in each pair. Because each trial contained six face pairs, this procedure yielded a vector of six predicted dissimilarities per trial.

During stimulus optimization, which was performed before the behavioral experiment, we replaced each trial-wise human response vector $\mathbf{y}_{t}$ with $\tilde{\mathbf{y}}^{m,l}_{t}$, the corresponding predicted dissimilarity vector produced by layer $l$ of the data-generating instance of model $m$. We used Pearson's correlation coefficient rather than Spearman's rank correlation to preserve differentiability during gradient-based stimulus optimization:
\begin{equation}
\psi_r \big(\tilde{\mathbf{y}}^{m,l}, \hat{\mathbf{y}}^{m',l'} \mid \xi\big)
= \frac{1}{T} \sum_{t=1}^{T} r\big(\tilde{\mathbf{y}}^{m,l}_{t}, \hat{\mathbf{y}}^{m',l'}_{t}\big).
\label{eq:model_performance_estimator_pearson}
\end{equation}

Thus, $\psi_r$ provides the concrete instantiation of the performance estimator $\psi$ in the stimulus-optimization objective (Eq.~\ref{eq:multi_layer_local_objective}), whereas $\psi_\rho$ was used for empirical model--behavior analysis.

\paragraph{Optimization procedure and implementation details}
To find stimulus sets for which the utility $U(\xi)$ is high, we randomly initialized the stimuli and used gradient ascent (Adam) to increase model discrimination as defined by Eq.~\ref{eq:multi_layer_global_objective}. The random initialization of each controversial stimulus set was seed-matched to one of the randomly sampled sets; therefore, differences between the paired random and controversial sets used in the behavioral experiment reflect the effect of the optimization procedure.

Condition-specific hyperparameters are summarized in Supplementary Table~\ref{tab:opt_hparams}. To increase robustness and simulate fixation variability across participants, we averaged utility across spatially jittered presentations before each gradient step. To ensure that the optimized stimuli remained face-like, we constrained the shape and texture latents during optimization for BFM stimuli (\ref{sec:supp_bfm_latent_constraint}); for StyleGAN3 stimuli, we added a face-log-probability term to the objective, computed using the YOLO5Face detector \cite{qi_yolo5face_2023} (see~\ref{sec:supp_face_probability_constraint} for how detection scores were transformed to probabilities). Because this term was defined as a log probability, low face probability sharply reduced the optimized objective, which discouraged the optimization from producing images that no longer appeared face-like. Full optimization implementation details (hardware allocation, convergence schedule, and the exact latent-space and face-probability formulations) are provided in~\ref{sec:supp_opt_details}.

\subsubsection{Model recovery simulation}
\label{sec:model_recovery}

We quantified how well each stimulus set supported recovery of the data-generating model in a simulation analysis. We treated each candidate model and each of its layers in turn as the data-generating process. In each simulation, layer $l$ of a held-out instance of the data-generating model $m$ was used to generate the ground-truth dissimilarity vector $\tilde{\mathbf{y}}^{m,l}$. We then compared this vector to the dissimilarity vectors $\hat{\mathbf{y}}^{m',l'}$ generated by the reference instances of all candidate models. Model recovery was counted as successful when the best-matching reference model was the data-generating model.

For noiseless simulations, model-recovery accuracy for stimulus set $\xi$ was defined as
\begin{equation}
A(\xi) = \sum_m p(m)  \sum_{l=1}^{L_m} p(l \mid m) \, \mathbbm{1}\Big[\max\limits_{l'} \psi_{\rho}\big(\tilde{\mathbf{y}}^{m,l},\hat{\mathbf{y}}^{m,l'} \mid \xi \big) > \max\limits_{m' \neq m}\max\limits_{l'}\psi_{\rho}\big(\tilde{\mathbf{y}}^{m,l},\hat{\mathbf{y}}^{m',l'} \mid \xi \big)\Big],
\label{eq:model_recovery_objective}
\end{equation}
where $p(m)$ and $p(l \mid m)$ were uniform. Thus, model recovery was successful only when the simulated dissimilarity vector was more strongly correlated with at least one layer of the reference instance of the data-generating model than with any layer of any alternative model. Averaging over $m$ and $l$ yielded the expected probability of recovering the correct model.

To simulate measurement noise, we added independent zero-mean Gaussian noise after normalizing the dissimilarities. Specifically, for each data-generating model and layer, we computed the standard deviation of the corresponding noiseless dissimilarities across face pairs within each trial of a randomly sampled stimulus set, and then averaged these standard deviation estimates across trials. We divided each ground-truth dissimilarity vector by the resulting normalization scale before adding noise. Thus, the noise level was expressed in normalized dissimilarity units. For each stimulus set and noise level, we estimated $A(\xi)$ using 1000 random draws per model and layer.

\begin{figure}[t]
\centering
\makebox[\linewidth][c]{\includegraphics[]{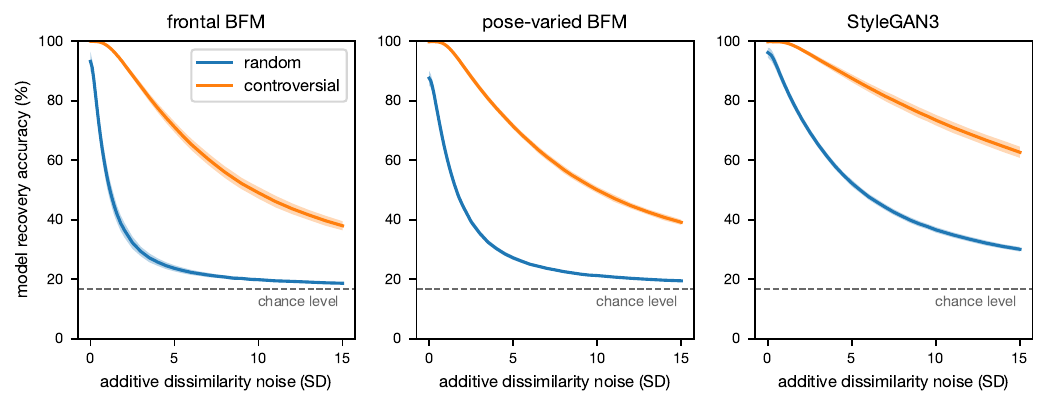}}
\caption{\textbf{Simulated model recovery accuracy across three stimulus families and two sampling strategies}. Columns show stimulus families, and colors show sampling strategies. To simulate observed dissimilarity vectors, we held out one instance of each model and treated each model and layer in turn as the data-generating process. These simulated observed dissimilarity vectors were compared against the dissimilarity vectors predicted by the reference instances used during stimulus optimization (Methods~\ref{sec:controversial_face_synthesis}). Measurement noise was introduced by adding independent Gaussian noise to each normalized dissimilarity (x-axis), with $\sigma = 1$ corresponding to the standard deviation of that representation's noiseless dissimilarities in an independently sampled random stimulus set. The y-axis shows the proportion of simulations in which the data-generating model was correctly identified, pooled across data-generating models and layers, and averaged across seeds. Shaded regions indicate standard deviation across seeds.}
\label{fig:model_recovery_acc}
\end{figure}

The simulated model-recovery curves in Fig.~\ref{fig:model_recovery_acc} therefore show, as a function of measurement noise, the proportion of simulations in which the data-generating model was recovered, irrespective of which layer generated the dissimilarities. As expected, the controversial stimulus sets yielded higher model-recovery accuracy than the randomly sampled sets across all stimulus families.

\subsubsection{Stimulus set selection}
For each controversial-face condition, we generated 36 candidate stimulus sets. We retained 12 sets for the behavioral experiments based on model-recovery accuracy, after excluding sets with visible generation artifacts or otherwise unnatural face images. The same retained random-initialization seeds were used to generate the corresponding randomly sampled face stimulus sets.

\subsection{Behavioral arrangement task}
\label{sec:behavioral_task}
We collected face-dissimilarity judgments from 864 participants (18--60 yr; M=36.58, SD=10.28; 435 self-identified as female) in a large-scale, between-subjects design through \url{prolific.co}. All participants self-reported normal or corrected-to-normal vision and provided informed consent. This research was approved by the Columbia University Institutional Review Board (protocol AAAR9520).

Each participant completed 24 trials (12 unique trials, each repeated once, in randomized order) drawn from one of 72 stimulus sets: 3 stimulus families (frontal BFM, pose-varied BFM, StyleGAN3) $\times$ 2 sampling strategies (randomly sampled, controversial) $\times$ 12 seeds. In each trial, participants arranged six face pairs according to their perceived dissimilarity along a vertical dissimilarity axis, anchored at ``same person'' and ``most dissimilar face pair in this trial'' (\figurename~\ref{fig:trial_illustration}a). The task was implemented by the Meadows web platform (\url{meadows-research.com}).

After completing a practice pair-arrangement trial, participants viewed an annotated illustration of the task and were instructed to arrange each face pair along the vertical axis according to perceived within-pair dissimilarity, placing the most dissimilar pair at the top and, when present, same-identity pairs at the bottom. For StyleGAN3 stimuli, participants were additionally instructed to focus on the faces rather than the background. The exact participant-facing instructions are provided in~\ref{sec:supp_task_details}. To ensure data quality, we excluded participants who failed any of three prespecified screening criteria (\ref{sec:supp_subject_screening}). We continued to collect data until each stimulus set was judged by 12 eligible participants. Participants were compensated at an average hourly rate of \$15.28; additional details of compensation and payment for excluded participants are provided in~\ref{sec:supp_compensation}.

\subsection{Model--behavior comparison and statistical inference}
\label{sec:model_behavior_comparison}
\subsubsection{Trial-level model--behavior comparison}
In each trial, participants arranged six face pairs along a vertical dissimilarity axis to indicate their perceived within-pair dissimilarities (Methods~\ref{sec:behavioral_task}). We used the vertical coordinates of the placed face pairs as the participants' dissimilarity responses. The horizontal coordinates were ignored, as the horizontal dimension was provided only to allow participants to place multiple pairs at the same vertical position.

To compute model performance, we first averaged the $y$ coordinates of each participant's responses across the two repetitions of the same trial and then rank-transformed the averaged responses within each trial. We applied the same rank transformation to the model-predicted dissimilarities and quantified model performance using $\psi_\rho$ as defined in Eq.~\ref{eq:model_performance_estimator_rho}. Ties were handled using the analytical random-among-equals tie-breaking estimate (Eq.~14 in the original formulation \cite{schutt_statistical_2023}).

\subsubsection{Noise ceiling estimates and leave-one-subject-out model}
\label{sec:noise_ceiling}

Models made the same ranking predictions for all participants who viewed the same stimulus set, but participants varied in their responses. For this reason, even an ideal model could not achieve perfect model--behavior agreement. We therefore quantified upper and lower bounds on model performance achievable given inter-subject variability (i.e., the noise ceiling; \cite{nili_toolbox_2014}).

Mathematically, the optimal rank vector $\mathbf{r}^*\in\mathbb{N}^d$ that maximizes the mean correlation with a set of rank vectors $\mathbf{r}_i\in\mathbb{N}^d$ for $i=1,\dots,n$ is:
\begin{align*}
    \mathbf{r}^*=\operatorname{ranks}\left(\frac{1}{n}\sum_{i=1}^n\mathbf{r}_i\right).
\end{align*}

We obtained an upper bound on the noise ceiling by rank-transforming the six pairwise judgments within each trial, averaging these ranks across all participants who viewed the same stimulus set, and rank-transforming the resulting averages within each trial. The resulting predictions were compared with the responses of each of these participants using the same model-performance estimator used for the model--human comparisons, $\psi_{\rho}$. After applying this procedure to all participants across all stimulus sets, the resulting estimates were averaged across all participants.

To obtain a lower bound on the noise ceiling, we used the same procedure but excluded the held-out participant's own judgments from the average rank vector used to predict their responses.

\subsubsection{Within-condition mixed-effects inference for model comparison}
\label{sec:statistical_inference_within}
We formally compared computational models within each stimulus condition (defined by the stimulus family and sampling strategy) using statistical inference. Because our goal was to compare models rather than specific layers, we estimated each model's best-layer performance using leave-one-subject-out cross-validation. Specifically, for model $m$ and held-out participant $p$, we selected the layer $l$ that maximized the average trial-wise Spearman correlation with the responses of the remaining participants. We then quantified the performance of model $m$ on each trial $t$ as the Spearman's correlation between the model-predicted dissimilarities, $\hat{\mathbf{y}}_t^{m,l}$, and participant $p$'s face-pair judgments, $\mathbf{y}^p_t$. This process yielded a single correlation coefficient per seed, model, participant, and trial.

Correlation coefficients were clipped to $[-1 + 10^{-6}, 1 - 10^{-6}]$ before Fisher-$z$ transformation to avoid infinite values. We then Fisher-$z$ transformed these clipped correlation coefficients and analyzed them separately for each stimulus condition using linear mixed-effects models implemented in the \texttt{lme4} package in R \cite{bates_fitting_2015, kuznetsova_lmertest_2017}. The model was specified as:
\begin{verbatim}
    z ~ model + (1|seed:subj_id) + (1|seed:trial_id)
\end{verbatim}
where \textit{model} was treated as a fixed effect representing the different models being compared. Random intercepts were included for participants and trials, each nested within stimulus-set seed, to account for repeated measurements and within-seed dependence.

We considered more complex random-effects structures, including random intercepts for seed and random slopes for model, but these yielded convergence failures or unstable parameter estimates across different numerical optimizers. We therefore selected the more parsimonious random-intercepts model above, which showed stable convergence while capturing the key sources of variance in the experimental design.

After fitting the model for each condition, we computed estimated marginal means for each model and performed two-sided pairwise model comparisons using the \texttt{emmeans} package \cite{lenth_emmeans}. Degrees of freedom were approximated using the Satterthwaite method, and resulting $p$ values were adjusted for multiple comparisons using the false discovery rate (FDR) procedure \citep{Benjamini_Hochberg_1995, kuznetsova_lmertest_2017}. Full parameter estimates and statistical test results are reported in Supplementary Table~\ref{tab:model_comparison_cond}.

\subsubsection{Pooled mixed-effects inference across stimulus conditions}
\label{sec:statistical_inference_pooled}
We also performed a pooled analysis that combined data across all six conditions (three stimulus families $\times$ two sampling methods). Model performance and layer selection for each participant were defined exactly as in the within-condition analysis (Methods~\ref{sec:statistical_inference_within}). This yielded one Spearman's rank correlation coefficient for each combination of stimulus family, sampling strategy, seed, model, participant, and trial.

To jointly compare models while accounting for the hierarchical structure of the data, we Fisher-$z$ transformed all correlation coefficients and fit a linear mixed-effects model to the pooled dataset:
\begin{verbatim}
    z ~ model * stimulus_family * sampling +
    (1 | stimulus_family:sampling:seed) +
    (1 | stimulus_family:sampling:seed:subj_id) +
    (1 | stimulus_family:sampling:seed:trial_id)
\end{verbatim}

where \textit{model}, \textit{stimulus\_family}, and \textit{sampling} were treated as fixed effects, together with all interaction terms. This full factorial specification allows model differences to vary across conditions. Random intercepts were included for stimulus-set seed, participant nested within seed, and trial nested within seed, capturing shared variance among repeated measurements from the same stimulus set and within-seed dependence across participants and trials.

We compared this full fixed-effects model with reduced alternatives fit by maximum likelihood. The full model provided the best fit according to Akaike's information criterion (AIC; \citep{akaike1974}), indicating that interactions between model and condition were warranted.

To quantify overall model performance, we computed estimated marginal means (EMMs) for each model, marginalized over the six stimulus-family-by-sampling conditions with equal weights \cite{lenth_emmeans}. We then performed two-sided pairwise model comparisons using these marginal means, again using the Satterthwaite method to approximate degrees of freedom. Resulting $p$ values were adjusted for multiple comparisons using the false discovery rate (FDR) procedure \cite{Benjamini_Hochberg_1995}.

\section*{Data availability}
Stimuli, anonymized human behavioral judgment data, and model dissimilarity predictions are available on the Open Science Framework (\url{https://osf.io/bzx4e}).

\section*{Code availability}
Code for generating randomly sampled and controversial stimuli, analyzing the behavioral data, and reproducing the figures is available on GitHub (\url{https://github.com/kriegeskorte-lab/controversial-face}). Trained model checkpoints are available on Hugging Face (\url{https://huggingface.co/wenx-guo/controversial-face-model-checkpoints}).

\section*{Acknowledgments}
Research reported in this publication was supported in part by the National Institute of Neurological Disorders and Stroke of the National Institutes of Health under award number 4R01NS128897. The content is solely the responsibility of the authors and does not necessarily represent the official views of the National Institutes of Health. The computational components of this work were performed on the Columbia Zuckerman Institute Axon GPU cluster. 

\section*{Author contributions}
W.G., H.H.S., N.K., and T.G. conceptualized and designed the study. W.G. collected the data. W.G., N.K., and T.G. analyzed the data. W.G., H.H.S., K.M.J., K.R.S., N.K., and T.G. interpreted the results. W.G., N.K., and T.G. wrote the initial manuscript draft, and all authors contributed to manuscript revision and approved the final version.

\backmatter

\bibliography{sn-bibliography}

\clearpage
\section*{Supplementary Information}

\setcounter{subsection}{0}
\renewcommand{\thesubsection}{Supplementary Methods \arabic{subsection}}
\renewcommand{\theHsubsection}{supplementary.\arabic{subsection}}

\setcounter{figure}{0}
\renewcommand{\figurename}{Supplementary Fig.}
\renewcommand{\thefigure}{\arabic{figure}}

\setcounter{table}{0}
\renewcommand{\tablename}{Supplementary Table}
\renewcommand{\thetable}{\arabic{table}}

\captionsetup[table]{
    labelfont={bf,large}
}

\subsection{Generating controversial face sets}
\label{sec:supp_opt_details}

\subsubsection{Optimization implementation details}
Stimuli were optimized on a GPU cluster using eight NVIDIA L40 GPUs. For each condition, a subset of GPUs was allocated to image synthesis (i.e. PyTorch3D rendering for BFM or StyleGAN3 generation) and a subset to model evaluation (144 face images evaluated on 12 networks: two instances $\times$ six models). We computed the objective using the post-ReLU activations from all 16 weight layers (13 convolutional and 3 fully connected). We enabled gradient checkpointing to reduce memory.

We optimized the utility function using the Adam optimizer. At each step, we jittered the stimuli ten times using differentiable translation for the reference instance, while the data-generating instance always received centered inputs. This jittering procedure approximated the between-subject fixation variability. We averaged the utility over the ten jittered presentations and then took a gradient-ascent step to increase the global utility. Condition-specific choices for initial learning rate and moving average coefficients ($\beta_1, \beta_2$) for the Adam optimizer, jitter amplitude, latent space and bounds, and minimum GPU allocation are summarized in Supplementary Table~\ref{tab:opt_hparams}. The complete implementation of the stimulus-synthesis pipeline is available at \url{https://github.com/kriegeskorte-lab/controversial-face}.

\begin{table}[h!]
\centering
\caption{Optimization hyperparameters, latent space, and compute allocation by stimulus family.}
\label{tab:opt_hparams}
\setlength{\tabcolsep}{6pt} 

\begin{tabular}{@{} >{\raggedright\arraybackslash}p{1.5cm} c c c >{\centering\arraybackslash}p{2cm} >{\raggedright\arraybackslash}p{2.2cm} >{\raggedright\arraybackslash}p{5.2cm} @{}}
\toprule
\thead{\textbf{Stimulus}\\\textbf{Family}} & \thead{\textbf{Init.}\\\textbf{LR}} & \thead{\textbf{Adam Betas}\\($\beta_1, \beta_2$)} & \thead{\textbf{Jitter}\\\textbf{(px)}} & \thead{\textbf{Latent Space}} & \thead{\textbf{Constraint}} & \thead{\textbf{Minimum GPU allocation}\\\textbf{total \# (for generation / for model activations)}} \\
\midrule

\makecell[l]{BFM\\(frontal)} & 0.5 & 0.85, 0.995 & 5 & Shape/Texture ($\boldsymbol{\alpha}$) & $\|\boldsymbol{\alpha}\|_{\infty}\leq2$ & \makecell[l]{RTX 2080 Ti: 5 (2/3)\\
L40/A40: 2 (1/1)} \\
\addlinespace 

\makecell[l]{BFM\\(pose-varied)} & 0.1 & 0.92, 0.999 & 5 & Shape/Texture ($\boldsymbol{\alpha}$) & $\|\boldsymbol{\alpha}\|_{\infty}\leq2$ & \makecell[l]{RTX 2080 Ti: 5 (2/3)\\
L40/A40: 2 (1/1)} \\
\addlinespace

StyleGAN3 & 0.08 & 0.90, 0.999 & 8 & $W$ space & Image-space prob. (YOLO5Face) & L40/A40: 2 (1/1) \\

\bottomrule
\end{tabular}
\end{table}

\subsubsection{Convergence schedule}
The initial learning rate was halved upon convergence, defined as no increase in the mean objective over the most recent 50 iterations relative to the preceding 50. This reduction was applied at most twice. Optimization terminated at the third convergence step or at 1000 iterations, whichever occurred first.

\subsubsection{BFM latent space and $\ell_{\infty}$ norm constraints.}
\label{sec:supp_bfm_latent_constraint}
We optimized the shape and texture latent parameters in the BFM space. For pose-varied BFM stimuli, we used the same pre-defined pose configuration as in random sampling (Methods~\ref{sec:random_sampling}). To ensure BFM produced natural faces, we constrained the latent coefficients during optimization. Unconstrained gradient ascent rapidly drives the coefficients of the leading principal components to several standard deviations away from the origin (average face), yielding implausible and caricatured faces. For each latent dimension $i$, we optimized the unbounded variable $\alpha_i'$ and then mapped it to the bounded coefficient $\alpha_i=2 \tanh(\alpha_i')$ to render face images. This transformation constrained each coefficient $\alpha_i \in (-2,2)$ and functioned as a differentiable $\ell_{\infty}$ norm constraint.

\subsubsection{StyleGAN3 latent space and face probability constraint.}
\label{sec:supp_face_probability_constraint}
Although we used the $Z$ latent space to randomly sample faces, controversial optimization was performed in the intermediate $W$ space. In StyleGAN3, the initially sampled latent $\mathbf{z} \sim N(0,\mathbf{I})$ was mapped to $\mathbf{w}=f(\mathbf{z})$ by the generator's mapping network ($f: Z \rightarrow W$). We chose the $W$ space for optimization because it yields better-conditioned image mapping with more disentanglement \cite{karras_2019_cvpr}; we also ran a model metamer experiment inspired by prior model-metamer work \cite{feather_model_2023} to test different latent spaces, and $W$ showed the most stable convergence and produced metamers that were perceptually closest to the target faces, relative to $Z$ and the style space $S$ \cite{alaluf_stylespace_2022}.

\begin{figure}[H]
\centering
\makebox[\linewidth][c]{\includegraphics[]{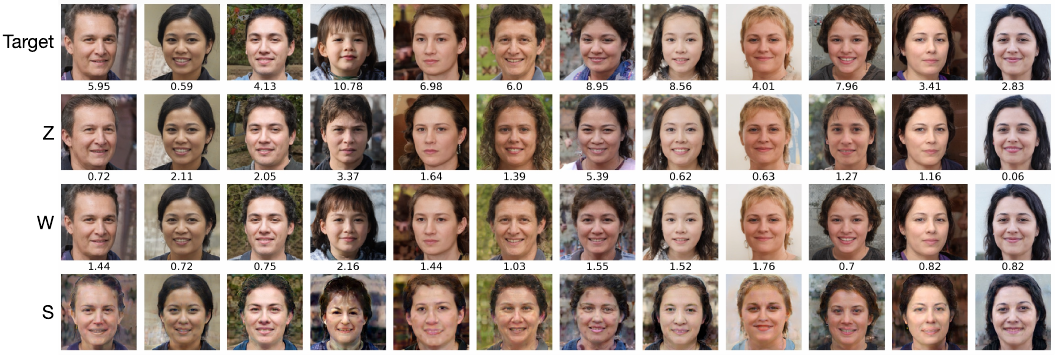}}
\caption{\textbf{Model metamer generation in the latent spaces of StyleGAN3.} Target faces were first generated by sampling latent codes from the StyleGAN3 $Z$ space. For each target image, we extracted activations from a late convolutional layer of the invRend-BFM model. We then initialized latent codes randomly in the $Z$, $W$, or $S$ space of StyleGAN3 and optimized them so that the resulting image elicited an activation pattern as close as possible to that of the target image. The synthesized images therefore constitute model metamers with respect to the selected invRend-BFM layer. The loss value shown above each synthesized face indicates the loss, with lower values reflecting a closer match to the target representation. We chose the $W$ space for controversial face optimization because it produced images that were perceptually closer to the targets and achieved lower final losses.}
\label{fig:model_metamer}
\end{figure}

Since the boundary between face and non-face images is not well-defined by a simple distribution in $W$ space (e.g. multivariate Gaussian), we added a face-probability constraint to the objective. For each generated image $x$ (downsampled to $32\times32$ pixels), YOLO5Face \cite{qi_yolo5face_2023}, a pretrained face detector, yielded multiple candidate boxes with confidence scores and classification probabilities. For each predicted box $i$, we computed a per-detection score $s_i=\text{conf}_i\times \text{cls}_i$, and pooled across detections with a differentiable smooth maximum operator:
\begin{equation}
    s_{\alpha}(x)=\frac{1}{\alpha}\log\sum_i\exp(\alpha s_i),\quad \alpha=20,
\end{equation}

which approaches $\max_i s_i(x)$ as $\alpha \rightarrow \infty$. We then converted $s_{\alpha}(x)$ into a calibrated face probability using a logistic classifier trained on face vs. non-face images ($4\times4$ grid-scrambled face images):
\begin{equation}
    p_{\text{face}}(x)=\sigma\big(\gamma s_{\alpha}(x)+\beta\big),
\end{equation}
where $\gamma$ and $\beta$ were fit once and held fixed during optimization. The utility function was then augmented by the average log likelihood over the batch $\xi$:
\begin{equation}
    \tilde{U}(\xi)=U(\xi)+\frac{1}{|\xi|}\sum_{x\in\xi}\log p_{\text{face}}(x).
\end{equation}

\subsection{Behavioral arrangement task}
\subsubsection{Participant instructions}
\label{sec:supp_task_details}
Following a practice trial of object-pair arrangement, participants viewed an annotated illustration of the face-pair arrangement task, with the following instructions on the right side:
\begin{displayquote}
You will place each face pair vertically according to the similarity of the two faces.\\\\
In each trial, place the pair with the two most dissimilar faces at the top. In case there are pairs showing the same person, place them at the bottom (next to the "same person" label). Place each other pair vertically, such that its position reflects the dissimilarity of the two faces.\\\\
Horizontal space is provided to allow placing pairs at the same vertical position (same similarity) while keeping all faces visible.
\end{displayquote}
For StyleGAN3 stimuli, participants were additionally instructed to focus on the faces rather than the background:
\begin{displayquote}
When making judgments, focus on the faces themselves, not their background.
\end{displayquote}

\subsubsection{Screening and exclusion criteria}
\label{sec:supp_subject_screening}
To ensure data quality, we applied three exclusion criteria. Participants were excluded if they failed \emph{any} one of the following:
\begin{enumerate}
    \item \textbf{Practice trial.} Before the main task, participants arranged image pairs of everyday objects. We required that their arrangement satisfied all three ordinal dissimilarity constraints:
    \begin{enumerate}
        \item $(\text{dog}, \text{car}) > (\text{apple}, \text{orange})$
        \item $(\text{apple}, \text{orange}) > (\text{apple}, \text{apple})$
        \item $(\text{apple}, \text{orange}) > (\text{strawberry}, \text{strawberry})$.
    \end{enumerate}
    Here, ``$>$'' denotes a strictly greater judged dissimilarity of the pair on the vertical axis.
    \item \textbf{Within-subject reliability.} For each of the 12 trials, we computed the rank correlation between the two repetitions (Spearman’s $\rho$) and applied Spearman-Brown correction. Corrected reliability estimates were clipped to $[-1 + 10^{-6}, 1 - 10^{-6}]$ before Fisher-z transformation. We then tested whether the mean $z$ exceeded 0 using a one-sample $t$-test. Here, $z=0$ corresponds to no correlation between two repetitions. Participants were excluded if this test was not significant ($p \ge .05$), indicating that their responses were not reliably correlated across repetitions.
    \item \textbf{Minimum task duration.} Participants with a total task time $<8$ minutes were excluded.
\end{enumerate}

\subsubsection{Participant compensation}
\label{sec:supp_compensation}
We paid \$6438 in total to $1073$ participants, including $209$ disqualified participants who failed one screening criterion. $44$ participants failed more than one screening criterion and were not paid. The hourly payment was \$15.28, on average.%

\clearpage
\begin{table}[h!]
\centering
\caption{Summary of model training procedures and final validation performance. The models were trained using PyTorch Lightning \cite{Falcon_PyTorch_Lightning_2019} on four NVIDIA GeForce RTX 2080 Ti GPUs, with dropout enabled during training phases only. All SGD optimizers used a momentum of 0.9. All Adam optimizers used $\beta_1=0.9$, $\beta_2=0.999$, and $\epsilon=10^{-4}$ unless otherwise specified. For autoencoders, metrics were computed over the full validation sets on $[0,1]$-scaled pixels using posterior-mean reconstructions and averaged across two instances; mean-image baseline RMSEs were 0.200 (VGGFace2) and 0.214 (BFM).}
\label{tab:model_training_summary}
\setlength{\tabcolsep}{4pt} 

\newcolumntype{L}[1]{>{\raggedright\arraybackslash}p{#1}}
\begin{tabularx}{\textwidth}{@{} >{\raggedright\arraybackslash}X l c c L{2.2cm} c c c @{}}
\toprule
\thead[l]{\textbf{Model Objective}} &
\thead[l]{\textbf{Dataset}} &
\thead[l]{\textbf{Opt.}} &
\thead[l]{\textbf{LR}} &
\thead[l]{\textbf{LR Schedule}} &
\thead[l]{\textbf{Batch}} &
\thead[l]{\textbf{Weight}\\\textbf{Decay}} &
\thead[l]{\textbf{Epochs}} \\
\midrule

\multicolumn{8}{@{}l}{\textbf{Supervised Classification}} \\
\addlinespace
Object Recognition & ImageNet & SGD & 0.1 & 10$\times$ drop / 30 ep. & 1024 & 1e-4 & 90 \\
\multicolumn{8}{l}{\hspace{1em}\textit{Validation Performance:} 62.4\% Top-1} \\
\cmidrule(lr){1-8}

Face Identification & VGGFace2 & SGD & 0.01 & 10$\times$ drop / 10 ep. & 1024 & 1e-4 & 30 \\
\multicolumn{8}{l}{\hspace{1em}\textit{Validation Performance:} 95.7\% Top-1} \\
\cmidrule(lr){1-8}

Face Identification & BFM & SGD & 0.01 & 10$\times$ drop / 10 ep. & 512 & 1e-4 & 30 \\
\multicolumn{8}{l}{\hspace{1em}\textit{Validation Performance:} 99.95\% Top-1} \\
\specialrule{1pt}{1pt}{1pt}

\multicolumn{8}{@{}l}{\textbf{Unsupervised Autoencoding}} \\
\addlinespace
Autoencoding & VGGFace2 & Adam & 5e-4 & 10$\times$ drop at 20 ep. & 256 & 1e-5 & 30 \\
\multicolumn{8}{l}{\hspace{1em}\textit{Validation Performance:} 0.00141 MSE; 0.0376 RMSE} \\
\cmidrule(lr){1-8}

Autoencoding & BFM & Adam & 5e-4 & 10$\times$ drop at 20 ep. & 256 & 1e-5 & 30 \\
\multicolumn{8}{l}{\hspace{1em}\textit{Validation Performance:} 0.00128 MSE; 0.0358 RMSE} \\
\specialrule{1pt}{1pt}{1pt}

\multicolumn{8}{@{}l}{\textbf{Inverse Generative}} \\
\addlinespace
\makecell[l]{Inverse rendering} & BFM & Adam & 1e-4 & 10$\times$ drop / 40 ep. & 512 & 0 & 120 \\
\multicolumn{8}{l}{\hspace{1em}\textit{Validation Performance:} 2.31 NMSE} \\

\bottomrule
\end{tabularx}
\end{table}

\clearpage
\input{cond_pairwise_latex}

\clearpage
\label{sec:pooled_tab}
\input{pooled_pairwise_latex}

\end{document}

%% file: cond_pairwise_latex.tex
\begin{small}

\begin{longtable}{
  @{}
  p{0.34\textwidth}
  S[table-format=1.3]
  c
  S[table-format=1.3]
  S[table-format=2.3]
  c
  c
  @{}
}

\caption{\textbf{Linear mixed-effects model estimates: condition-specific linear models.} The observation unit for these analyses was the trial-wise model--behavior correlation, with non-independence across observations accounted for by the random-effects structure. Test statistics were evaluated using Satterthwaite’s approximation to the denominator degrees of freedom, as implemented in \texttt{lmerTest} \cite{bates_fitting_2015, kuznetsova_lmertest_2017} and \texttt{emmeans} \citep{lenth_emmeans}. This approximation accounts for uncertainty in the estimated random-effect variance components and therefore yields contrast-specific effective degrees of freedom. Because the resulting degrees of freedom were large, the corresponding t-distributions were numerically close to a standard normal distribution.}
\label{tab:model_comparison_cond} \\
\toprule
\textbf{Comparison pair} &
{$\boldsymbol{\Delta} z$} &
{$\boldsymbol{\Delta} z$ 95\% CI} &
{\textbf{SE}} &
{\textbf{$t$-value}} &
{\textbf{$p$-value}} &
{\textbf{$p$-value (corrected)}} \\
\midrule
\endfirsthead

\multicolumn{7}{c}%
{{\bfseries\tablename\ \thetable{} -- continued from previous page}} \\
\toprule
\textbf{Comparison pair} &
{$\boldsymbol{\Delta} z$} &
{$\boldsymbol{\Delta} z$ 95\% CI} &
{\textbf{SE}} &
{\textbf{$t$-value}} &
{\textbf{$p$-value}} &
{\textbf{$p$-value (corrected)}} \\
\midrule
\endhead

\midrule
\multicolumn{7}{r}{{Continued on next page}} \\
\endfoot

\bottomrule
\endlastfoot

\multicolumn{7}{@{}l}{\textbf{Stimulus family: BFM -- Sampling strategy: controversial}} \\
\addlinespace
noise ceiling - invRend-BFM & 0.246 & \([0.183, 0.309]\) & 0.032 & 7.603 & \(3.60 \times 10^{-14}\) & \(7.56 \times 10^{-14}\) \\
noise ceiling - objCat-ImageNet & 0.275 & \([0.211, 0.338]\) & 0.032 & 8.491 & \(2.86 \times 10^{-17}\) & \(7.50 \times 10^{-17}\) \\
noise ceiling - autoenc-VGGFace2 & 0.277 & \([0.214, 0.341]\) & 0.032 & 8.579 & \(1.36 \times 10^{-17}\) & \(4.07 \times 10^{-17}\) \\
noise ceiling - faceID-VGGFace2 & 0.292 & \([0.229, 0.356]\) & 0.032 & 9.035 & \(2.51 \times 10^{-19}\) & \(1.32 \times 10^{-18}\) \\
noise ceiling - faceID-BFM & 0.377 & \([0.314, 0.441]\) & 0.032 & 11.664 & \(6.29 \times 10^{-31}\) & \(6.61 \times 10^{-30}\) \\
noise ceiling - autoenc-BFM & 0.501 & \([0.438, 0.565]\) & 0.032 & 15.493 & \(1.36 \times 10^{-52}\) & \(2.87 \times 10^{-51}\) \\
invRend-BFM - objCat-ImageNet & 0.029 & \([-0.021, 0.079]\) & 0.026 & 1.124 & 0.261 & 0.305 \\
invRend-BFM - autoenc-VGGFace2 & 0.032 & \([-0.019, 0.082]\) & 0.026 & 1.235 & 0.217 & 0.268 \\
invRend-BFM - faceID-VGGFace2 & 0.046 & \([-0.004, 0.096]\) & 0.026 & 1.812 & 0.0700 & 0.0919 \\
invRend-BFM - faceID-BFM & 0.131 & \([0.081, 0.181]\) & 0.026 & 5.137 & \(2.84 \times 10^{-7}\) & \(5.42 \times 10^{-7}\) \\
invRend-BFM - autoenc-BFM & 0.255 & \([0.205, 0.305]\) & 0.026 & 9.981 & \(2.30 \times 10^{-23}\) & \(1.61 \times 10^{-22}\) \\
objCat-ImageNet - autoenc-VGGFace2 & 0.003 & \([-0.047, 0.053]\) & 0.026 & 0.111 & 0.912 & 0.912 \\
objCat-ImageNet - faceID-VGGFace2 & 0.018 & \([-0.033, 0.068]\) & 0.026 & 0.688 & 0.491 & 0.543 \\
objCat-ImageNet - faceID-BFM & 0.103 & \([0.052, 0.153]\) & 0.026 & 4.013 & \(6.03 \times 10^{-5}\) & \(9.74 \times 10^{-5}\) \\
objCat-ImageNet - autoenc-BFM & 0.226 & \([0.176, 0.277]\) & 0.026 & 8.857 & \(9.42 \times 10^{-19}\) & \(3.96 \times 10^{-18}\) \\
autoenc-VGGFace2 - faceID-VGGFace2 & 0.015 & \([-0.035, 0.065]\) & 0.026 & 0.577 & 0.564 & 0.592 \\
autoenc-VGGFace2 - faceID-BFM & 0.100 & \([0.050, 0.150]\) & 0.026 & 3.902 & \(9.60 \times 10^{-5}\) & \(1.44 \times 10^{-4}\) \\
autoenc-VGGFace2 - autoenc-BFM & 0.224 & \([0.174, 0.274]\) & 0.026 & 8.746 & \(2.52 \times 10^{-18}\) & \(8.82 \times 10^{-18}\) \\
faceID-VGGFace2 - faceID-BFM & 0.085 & \([0.035, 0.135]\) & 0.026 & 3.325 & \(8.87 \times 10^{-4}\) & 0.00124 \\
faceID-VGGFace2 - autoenc-BFM & 0.209 & \([0.159, 0.259]\) & 0.026 & 8.169 & \(3.44 \times 10^{-16}\) & \(8.02 \times 10^{-16}\) \\
faceID-BFM - autoenc-BFM & 0.124 & \([0.074, 0.174]\) & 0.026 & 4.844 & \(1.29 \times 10^{-6}\) & \(2.26 \times 10^{-6}\) \\
\specialrule{1pt}{1pt}{1pt}
\multicolumn{7}{@{}l}{\textbf{Stimulus family: BFM -- Sampling strategy: random}} \\
\addlinespace
noise ceiling - autoenc-VGGFace2 & 0.140 & \([0.068, 0.211]\) & 0.037 & 3.827 & \(1.31 \times 10^{-4}\) & \(4.60 \times 10^{-4}\) \\
noise ceiling - faceID-VGGFace2 & 0.147 & \([0.075, 0.218]\) & 0.037 & 4.011 & \(6.14 \times 10^{-5}\) & \(2.58 \times 10^{-4}\) \\
noise ceiling - invRend-BFM & 0.158 & \([0.087, 0.230]\) & 0.037 & 4.329 & \(1.52 \times 10^{-5}\) & \(8.00 \times 10^{-5}\) \\
noise ceiling - faceID-BFM & 0.185 & \([0.114, 0.257]\) & 0.037 & 5.076 & \(3.99 \times 10^{-7}\) & \(2.79 \times 10^{-6}\) \\
noise ceiling - autoenc-BFM & 0.189 & \([0.117, 0.260]\) & 0.037 & 5.167 & \(2.47 \times 10^{-7}\) & \(2.60 \times 10^{-6}\) \\
noise ceiling - objCat-ImageNet & 0.202 & \([0.130, 0.273]\) & 0.037 & 5.518 & \(3.59 \times 10^{-8}\) & \(7.55 \times 10^{-7}\) \\
autoenc-VGGFace2 - faceID-VGGFace2 & 0.007 & \([-0.049, 0.063]\) & 0.029 & 0.234 & 0.815 & 0.855 \\
autoenc-VGGFace2 - invRend-BFM & 0.018 & \([-0.038, 0.075]\) & 0.029 & 0.641 & 0.522 & 0.685 \\
autoenc-VGGFace2 - faceID-BFM & 0.046 & \([-0.011, 0.102]\) & 0.029 & 1.593 & 0.111 & 0.233 \\
autoenc-VGGFace2 - autoenc-BFM & 0.049 & \([-0.007, 0.105]\) & 0.029 & 1.708 & 0.0876 & 0.204 \\
autoenc-VGGFace2 - objCat-ImageNet & 0.062 & \([0.006, 0.118]\) & 0.029 & 2.157 & 0.0311 & 0.0932 \\
faceID-VGGFace2 - invRend-BFM & 0.012 & \([-0.045, 0.068]\) & 0.029 & 0.406 & 0.685 & 0.757 \\
faceID-VGGFace2 - faceID-BFM & 0.039 & \([-0.017, 0.095]\) & 0.029 & 1.359 & 0.174 & 0.282 \\
faceID-VGGFace2 - autoenc-BFM & 0.042 & \([-0.014, 0.098]\) & 0.029 & 1.474 & 0.140 & 0.246 \\
faceID-VGGFace2 - objCat-ImageNet & 0.055 & \([-0.001, 0.111]\) & 0.029 & 1.922 & 0.0546 & 0.143 \\
invRend-BFM - faceID-BFM & 0.027 & \([-0.029, 0.083]\) & 0.029 & 0.952 & 0.341 & 0.477 \\
invRend-BFM - autoenc-BFM & 0.031 & \([-0.026, 0.087]\) & 0.029 & 1.068 & 0.286 & 0.428 \\
invRend-BFM - objCat-ImageNet & 0.043 & \([-0.013, 0.100]\) & 0.029 & 1.516 & 0.130 & 0.246 \\
faceID-BFM - autoenc-BFM & 0.003 & \([-0.053, 0.059]\) & 0.029 & 0.115 & 0.908 & 0.908 \\
faceID-BFM - objCat-ImageNet & 0.016 & \([-0.040, 0.072]\) & 0.029 & 0.564 & 0.573 & 0.708 \\
autoenc-BFM - objCat-ImageNet & 0.013 & \([-0.043, 0.069]\) & 0.029 & 0.448 & 0.654 & 0.757 \\
\specialrule{1pt}{1pt}{1pt}
\multicolumn{7}{@{}l}{\textbf{Stimulus family: pose-varied BFM -- Sampling strategy: controversial}} \\
\addlinespace
noise ceiling - autoenc-VGGFace2 & 0.471 & \([0.406, 0.536]\) & 0.033 & 14.228 & \(3.03 \times 10^{-45}\) & \(1.06 \times 10^{-44}\) \\
noise ceiling - invRend-BFM & 0.476 & \([0.411, 0.541]\) & 0.033 & 14.373 & \(4.05 \times 10^{-46}\) & \(1.70 \times 10^{-45}\) \\
noise ceiling - faceID-BFM & 0.534 & \([0.469, 0.599]\) & 0.033 & 16.120 & \(2.49 \times 10^{-57}\) & \(1.31 \times 10^{-56}\) \\
noise ceiling - objCat-ImageNet & 0.609 & \([0.544, 0.674]\) & 0.033 & 18.384 & \(1.39 \times 10^{-73}\) & \(9.71 \times 10^{-73}\) \\
noise ceiling - faceID-VGGFace2 & 0.723 & \([0.658, 0.788]\) & 0.033 & 21.824 & \(7.07 \times 10^{-102}\) & \(7.43 \times 10^{-101}\) \\
noise ceiling - autoenc-BFM & 0.791 & \([0.726, 0.856]\) & 0.033 & 23.880 & \(9.18 \times 10^{-121}\) & \(1.93 \times 10^{-119}\) \\
autoenc-VGGFace2 - invRend-BFM & 0.005 & \([-0.048, 0.058]\) & 0.027 & 0.178 & 0.859 & 0.859 \\
autoenc-VGGFace2 - faceID-BFM & 0.063 & \([0.010, 0.115]\) & 0.027 & 2.328 & 0.0199 & 0.0220 \\
autoenc-VGGFace2 - objCat-ImageNet & 0.138 & \([0.085, 0.190]\) & 0.027 & 5.113 & \(3.22 \times 10^{-7}\) & \(4.83 \times 10^{-7}\) \\
autoenc-VGGFace2 - faceID-VGGFace2 & 0.252 & \([0.199, 0.304]\) & 0.027 & 9.345 & \(1.09 \times 10^{-20}\) & \(2.29 \times 10^{-20}\) \\
autoenc-VGGFace2 - autoenc-BFM & 0.320 & \([0.267, 0.372]\) & 0.027 & 11.874 & \(2.50 \times 10^{-32}\) & \(7.50 \times 10^{-32}\) \\
invRend-BFM - faceID-BFM & 0.058 & \([0.005, 0.111]\) & 0.027 & 2.150 & 0.0316 & 0.0331 \\
invRend-BFM - objCat-ImageNet & 0.133 & \([0.080, 0.186]\) & 0.027 & 4.935 & \(8.12 \times 10^{-7}\) & \(1.14 \times 10^{-6}\) \\
invRend-BFM - faceID-VGGFace2 & 0.247 & \([0.194, 0.300]\) & 0.027 & 9.167 & \(5.70 \times 10^{-20}\) & \(1.09 \times 10^{-19}\) \\
invRend-BFM - autoenc-BFM & 0.315 & \([0.262, 0.368]\) & 0.027 & 11.696 & \(2.01 \times 10^{-31}\) & \(5.29 \times 10^{-31}\) \\
faceID-BFM - objCat-ImageNet & 0.075 & \([0.022, 0.128]\) & 0.027 & 2.785 & 0.00536 & 0.00662 \\
faceID-BFM - faceID-VGGFace2 & 0.189 & \([0.136, 0.242]\) & 0.027 & 7.016 & \(2.41 \times 10^{-12}\) & \(4.21 \times 10^{-12}\) \\
faceID-BFM - autoenc-BFM & 0.257 & \([0.204, 0.310]\) & 0.027 & 9.545 & \(1.63 \times 10^{-21}\) & \(3.79 \times 10^{-21}\) \\
objCat-ImageNet - faceID-VGGFace2 & 0.114 & \([0.061, 0.167]\) & 0.027 & 4.231 & \(2.34 \times 10^{-5}\) & \(3.07 \times 10^{-5}\) \\
objCat-ImageNet - autoenc-BFM & 0.182 & \([0.129, 0.235]\) & 0.027 & 6.761 & \(1.44 \times 10^{-11}\) & \(2.33 \times 10^{-11}\) \\
faceID-VGGFace2 - autoenc-BFM & 0.068 & \([0.015, 0.121]\) & 0.027 & 2.529 & 0.0114 & 0.0134 \\
\specialrule{1pt}{1pt}{1pt}
\multicolumn{7}{@{}l}{\textbf{Stimulus family: pose-varied BFM -- Sampling strategy: random}} \\
\addlinespace
noise ceiling - invRend-BFM & 0.501 & \([0.440, 0.561]\) & 0.031 & 16.204 & \(9.47 \times 10^{-58}\) & \(3.32 \times 10^{-57}\) \\
noise ceiling - faceID-BFM & 0.502 & \([0.441, 0.563]\) & 0.031 & 16.243 & \(5.22 \times 10^{-58}\) & \(2.19 \times 10^{-57}\) \\
noise ceiling - autoenc-VGGFace2 & 0.507 & \([0.447, 0.568]\) & 0.031 & 16.410 & \(3.78 \times 10^{-59}\) & \(1.99 \times 10^{-58}\) \\
noise ceiling - objCat-ImageNet & 0.535 & \([0.474, 0.595]\) & 0.031 & 17.304 & \(2.14 \times 10^{-65}\) & \(1.50 \times 10^{-64}\) \\
noise ceiling - autoenc-BFM & 0.568 & \([0.507, 0.629]\) & 0.031 & 18.379 & \(2.64 \times 10^{-73}\) & \(2.77 \times 10^{-72}\) \\
noise ceiling - faceID-VGGFace2 & 0.601 & \([0.540, 0.661]\) & 0.031 & 19.435 & \(1.79 \times 10^{-81}\) & \(3.76 \times 10^{-80}\) \\
invRend-BFM - faceID-BFM & 0.001 & \([-0.050, 0.052]\) & 0.026 & 0.046 & 0.964 & 0.964 \\
invRend-BFM - autoenc-VGGFace2 & 0.006 & \([-0.044, 0.057]\) & 0.026 & 0.246 & 0.806 & 0.883 \\
invRend-BFM - objCat-ImageNet & 0.034 & \([-0.017, 0.085]\) & 0.026 & 1.312 & 0.189 & 0.257 \\
invRend-BFM - autoenc-BFM & 0.067 & \([0.016, 0.118]\) & 0.026 & 2.596 & 0.00945 & 0.0193 \\
invRend-BFM - faceID-VGGFace2 & 0.100 & \([0.049, 0.151]\) & 0.026 & 3.855 & \(1.16 \times 10^{-4}\) & \(3.49 \times 10^{-4}\) \\
faceID-BFM - autoenc-VGGFace2 & 0.005 & \([-0.046, 0.056]\) & 0.026 & 0.200 & 0.841 & 0.883 \\
faceID-BFM - objCat-ImageNet & 0.033 & \([-0.018, 0.084]\) & 0.026 & 1.267 & 0.205 & 0.257 \\
faceID-BFM - autoenc-BFM & 0.066 & \([0.015, 0.117]\) & 0.026 & 2.550 & 0.0108 & 0.0193 \\
faceID-BFM - faceID-VGGFace2 & 0.099 & \([0.048, 0.149]\) & 0.026 & 3.809 & \(1.40 \times 10^{-4}\) & \(3.67 \times 10^{-4}\) \\
autoenc-VGGFace2 - objCat-ImageNet & 0.028 & \([-0.023, 0.078]\) & 0.026 & 1.066 & 0.286 & 0.334 \\
autoenc-VGGFace2 - autoenc-BFM & 0.061 & \([0.010, 0.112]\) & 0.026 & 2.350 & 0.0188 & 0.0304 \\
autoenc-VGGFace2 - faceID-VGGFace2 & 0.093 & \([0.043, 0.144]\) & 0.026 & 3.609 & \(3.08 \times 10^{-4}\) & \(7.19 \times 10^{-4}\) \\
objCat-ImageNet - autoenc-BFM & 0.033 & \([-0.018, 0.084]\) & 0.026 & 1.283 & 0.199 & 0.257 \\
objCat-ImageNet - faceID-VGGFace2 & 0.066 & \([0.015, 0.117]\) & 0.026 & 2.543 & 0.0110 & 0.0193 \\
autoenc-BFM - faceID-VGGFace2 & 0.033 & \([-0.018, 0.083]\) & 0.026 & 1.259 & 0.208 & 0.257 \\
\specialrule{1pt}{1pt}{1pt}
\multicolumn{7}{@{}l}{\textbf{Stimulus family: StyleGAN3 -- Sampling strategy: controversial}} \\
\addlinespace
noise ceiling - objCat-ImageNet & 0.523 & \([0.460, 0.585]\) & 0.032 & 16.291 & \(4.59 \times 10^{-58}\) & \(1.61 \times 10^{-57}\) \\
noise ceiling - faceID-VGGFace2 & 0.572 & \([0.509, 0.635]\) & 0.032 & 17.831 & \(7.71 \times 10^{-69}\) & \(3.24 \times 10^{-68}\) \\
noise ceiling - faceID-BFM & 0.726 & \([0.663, 0.789]\) & 0.032 & 22.624 & \(1.42 \times 10^{-107}\) & \(7.47 \times 10^{-107}\) \\
noise ceiling - autoenc-BFM & 0.800 & \([0.737, 0.863]\) & 0.032 & 24.944 & \(5.81 \times 10^{-129}\) & \(4.07 \times 10^{-128}\) \\
noise ceiling - invRend-BFM & 0.813 & \([0.751, 0.876]\) & 0.032 & 25.359 & \(5.80 \times 10^{-133}\) & \(6.09 \times 10^{-132}\) \\
noise ceiling - autoenc-VGGFace2 & 0.837 & \([0.774, 0.900]\) & 0.032 & 26.093 & \(3.81 \times 10^{-140}\) & \(8.01 \times 10^{-139}\) \\
objCat-ImageNet - faceID-VGGFace2 & 0.049 & \([-0.003, 0.102]\) & 0.027 & 1.858 & 0.0632 & 0.0737 \\
objCat-ImageNet - faceID-BFM & 0.203 & \([0.151, 0.255]\) & 0.027 & 7.639 & \(2.36 \times 10^{-14}\) & \(3.81 \times 10^{-14}\) \\
objCat-ImageNet - autoenc-BFM & 0.278 & \([0.225, 0.330]\) & 0.027 & 10.438 & \(2.16 \times 10^{-25}\) & \(5.04 \times 10^{-25}\) \\
objCat-ImageNet - invRend-BFM & 0.291 & \([0.239, 0.343]\) & 0.027 & 10.939 & \(1.02 \times 10^{-27}\) & \(2.69 \times 10^{-27}\) \\
objCat-ImageNet - autoenc-VGGFace2 & 0.314 & \([0.262, 0.367]\) & 0.027 & 11.824 & \(4.48 \times 10^{-32}\) & \(1.34 \times 10^{-31}\) \\
faceID-VGGFace2 - faceID-BFM & 0.154 & \([0.102, 0.206]\) & 0.027 & 5.781 & \(7.63 \times 10^{-9}\) & \(1.14 \times 10^{-8}\) \\
faceID-VGGFace2 - autoenc-BFM & 0.228 & \([0.176, 0.280]\) & 0.027 & 8.580 & \(1.07 \times 10^{-17}\) & \(1.88 \times 10^{-17}\) \\
faceID-VGGFace2 - invRend-BFM & 0.241 & \([0.189, 0.294]\) & 0.027 & 9.081 & \(1.25 \times 10^{-19}\) & \(2.39 \times 10^{-19}\) \\
faceID-VGGFace2 - autoenc-VGGFace2 & 0.265 & \([0.213, 0.317]\) & 0.027 & 9.966 & \(2.67 \times 10^{-23}\) & \(5.60 \times 10^{-23}\) \\
faceID-BFM - autoenc-BFM & 0.074 & \([0.022, 0.127]\) & 0.027 & 2.799 & 0.00514 & 0.00635 \\
faceID-BFM - invRend-BFM & 0.088 & \([0.036, 0.140]\) & 0.027 & 3.300 & \(9.69 \times 10^{-4}\) & 0.00127 \\
faceID-BFM - autoenc-VGGFace2 & 0.111 & \([0.059, 0.163]\) & 0.027 & 4.185 & \(2.87 \times 10^{-5}\) & \(4.02 \times 10^{-5}\) \\
autoenc-BFM - invRend-BFM & 0.013 & \([-0.039, 0.065]\) & 0.027 & 0.501 & 0.616 & 0.616 \\
autoenc-BFM - autoenc-VGGFace2 & 0.037 & \([-0.015, 0.089]\) & 0.027 & 1.386 & 0.166 & 0.183 \\
invRend-BFM - autoenc-VGGFace2 & 0.024 & \([-0.029, 0.076]\) & 0.027 & 0.885 & 0.376 & 0.395 \\
\specialrule{1pt}{1pt}{1pt}
\multicolumn{7}{@{}l}{\textbf{Stimulus family: StyleGAN3 -- Sampling strategy: random}} \\
\addlinespace
noise ceiling - faceID-VGGFace2 & 0.420 & \([0.353, 0.487]\) & 0.034 & 12.268 & \(3.03 \times 10^{-34}\) & \(1.06 \times 10^{-33}\) \\
noise ceiling - objCat-ImageNet & 0.534 & \([0.467, 0.601]\) & 0.034 & 15.608 & \(5.18 \times 10^{-54}\) & \(2.18 \times 10^{-53}\) \\
noise ceiling - invRend-BFM & 0.593 & \([0.526, 0.660]\) & 0.034 & 17.324 & \(7.23 \times 10^{-66}\) & \(3.80 \times 10^{-65}\) \\
noise ceiling - faceID-BFM & 0.697 & \([0.630, 0.764]\) & 0.034 & 20.372 & \(1.15 \times 10^{-89}\) & \(8.07 \times 10^{-89}\) \\
noise ceiling - autoenc-VGGFace2 & 0.707 & \([0.640, 0.774]\) & 0.034 & 20.665 & \(3.86 \times 10^{-92}\) & \(4.05 \times 10^{-91}\) \\
noise ceiling - autoenc-BFM & 0.716 & \([0.649, 0.783]\) & 0.034 & 20.927 & \(2.27 \times 10^{-94}\) & \(4.77 \times 10^{-93}\) \\
faceID-VGGFace2 - objCat-ImageNet & 0.114 & \([0.060, 0.169]\) & 0.028 & 4.105 & \(4.08 \times 10^{-5}\) & \(5.35 \times 10^{-5}\) \\
faceID-VGGFace2 - invRend-BFM & 0.173 & \([0.118, 0.228]\) & 0.028 & 6.213 & \(5.37 \times 10^{-10}\) & \(9.41 \times 10^{-10}\) \\
faceID-VGGFace2 - faceID-BFM & 0.277 & \([0.223, 0.332]\) & 0.028 & 9.958 & \(2.90 \times 10^{-23}\) & \(6.77 \times 10^{-23}\) \\
faceID-VGGFace2 - autoenc-VGGFace2 & 0.287 & \([0.233, 0.342]\) & 0.028 & 10.319 & \(7.45 \times 10^{-25}\) & \(1.95 \times 10^{-24}\) \\
faceID-VGGFace2 - autoenc-BFM & 0.296 & \([0.242, 0.351]\) & 0.028 & 10.640 & \(2.57 \times 10^{-26}\) & \(7.71 \times 10^{-26}\) \\
objCat-ImageNet - invRend-BFM & 0.059 & \([0.004, 0.113]\) & 0.028 & 2.109 & 0.0350 & 0.0408 \\
objCat-ImageNet - faceID-BFM & 0.163 & \([0.108, 0.217]\) & 0.028 & 5.853 & \(4.96 \times 10^{-9}\) & \(8.01 \times 10^{-9}\) \\
objCat-ImageNet - autoenc-VGGFace2 & 0.173 & \([0.118, 0.228]\) & 0.028 & 6.214 & \(5.34 \times 10^{-10}\) & \(9.41 \times 10^{-10}\) \\
objCat-ImageNet - autoenc-BFM & 0.182 & \([0.127, 0.236]\) & 0.028 & 6.536 & \(6.61 \times 10^{-11}\) & \(1.39 \times 10^{-10}\) \\
invRend-BFM - faceID-BFM & 0.104 & \([0.050, 0.159]\) & 0.028 & 3.745 & \(1.82 \times 10^{-4}\) & \(2.24 \times 10^{-4}\) \\
invRend-BFM - autoenc-VGGFace2 & 0.114 & \([0.060, 0.169]\) & 0.028 & 4.106 & \(4.06 \times 10^{-5}\) & \(5.35 \times 10^{-5}\) \\
invRend-BFM - autoenc-BFM & 0.123 & \([0.069, 0.178]\) & 0.028 & 4.427 & \(9.64 \times 10^{-6}\) & \(1.45 \times 10^{-5}\) \\
faceID-BFM - autoenc-VGGFace2 & 0.010 & \([-0.045, 0.065]\) & 0.028 & 0.361 & 0.718 & 0.748 \\
faceID-BFM - autoenc-BFM & 0.019 & \([-0.036, 0.074]\) & 0.028 & 0.682 & 0.495 & 0.547 \\
autoenc-VGGFace2 - autoenc-BFM & 0.009 & \([-0.046, 0.064]\) & 0.028 & 0.321 & 0.748 & 0.748 \\

\end{longtable}

\end{small}

%% file: pooled_pairwise_latex.tex
\begin{small}

\begin{longtable}{
  @{}
  p{0.34\textwidth}
  S[table-format=1.3]
  c
  S[table-format=1.3]
  S[table-format=2.3]
  c
  c
  @{}
}

\caption{\textbf{Linear mixed-effects model estimates for condition-pooled model comparisons.} Test statistics were evaluated with Satterthwaite-approximated denominator degrees of freedom, which account for uncertainty in the estimated variance components and can differ across contrasts. Because the resulting degrees of freedom were large, the corresponding t-distributions were effectively normal.}
\label{tab:model_comparison_pooled} \\
\toprule
\textbf{Comparison pair} &
{$\boldsymbol{\Delta} z$} &
{$\boldsymbol{\Delta} z$ 95\% CI} &
{\textbf{SE}} &
{\textbf{$t$-value}} &
{\textbf{$p$-value}} &
{\textbf{$p$-value (corrected)}} \\
\midrule
\endfirsthead

\multicolumn{7}{c}%
{{\bfseries\tablename\ \thetable{} -- continued from previous page}} \\
\toprule
\textbf{Comparison pair} &
{$\boldsymbol{\Delta} z$} &
{$\boldsymbol{\Delta} z$ 95\% CI} &
{\textbf{SE}} &
{\textbf{$t$-value}} &
{\textbf{$p$-value}} &
{\textbf{$p$-value (corrected)}} \\
\midrule
\endhead

\midrule
\multicolumn{7}{r}{{Continued on next page}} \\
\endfoot

\bottomrule
\endlastfoot

noise ceiling - objCat-ImageNet & 0.443 & \([0.416, 0.471]\) & 0.014 & 31.927 & \(1.06 \times 10^{-197}\) & \(3.72 \times 10^{-197}\) \\
noise ceiling - faceID-VGGFace2 & 0.456 & \([0.429, 0.483]\) & 0.014 & 32.856 & \(6.84 \times 10^{-208}\) & \(2.87 \times 10^{-207}\) \\
noise ceiling - invRend-BFM & 0.462 & \([0.435, 0.489]\) & 0.014 & 33.254 & \(2.70 \times 10^{-212}\) & \(1.42 \times 10^{-211}\) \\
noise ceiling - autoenc-VGGFace2 & 0.487 & \([0.460, 0.514]\) & 0.014 & 35.086 & \(5.87 \times 10^{-233}\) & \(4.11 \times 10^{-232}\) \\
noise ceiling - faceID-BFM & 0.501 & \([0.474, 0.528]\) & 0.014 & 36.065 & \(3.16 \times 10^{-244}\) & \(3.32 \times 10^{-243}\) \\
noise ceiling - autoenc-BFM & 0.591 & \([0.564, 0.619]\) & 0.014 & 42.590 & \(1.19 \times 10^{-322}\) & \(2.49 \times 10^{-321}\) \\
objCat-ImageNet - faceID-VGGFace2 & 0.013 & \([-0.009, 0.034]\) & 0.011 & 1.175 & 0.240 & 0.252 \\
objCat-ImageNet - invRend-BFM & 0.018 & \([-0.003, 0.040]\) & 0.011 & 1.677 & 0.0936 & 0.109 \\
objCat-ImageNet - autoenc-VGGFace2 & 0.044 & \([0.022, 0.065]\) & 0.011 & 3.991 & \(6.59 \times 10^{-5}\) & \(9.89 \times 10^{-5}\) \\
objCat-ImageNet - faceID-BFM & 0.057 & \([0.036, 0.079]\) & 0.011 & 5.227 & \(1.72 \times 10^{-7}\) & \(3.02 \times 10^{-7}\) \\
objCat-ImageNet - autoenc-BFM & 0.148 & \([0.127, 0.170]\) & 0.011 & 13.470 & \(2.63 \times 10^{-41}\) & \(7.90 \times 10^{-41}\) \\
faceID-VGGFace2 - invRend-BFM & 0.006 & \([-0.016, 0.027]\) & 0.011 & 0.502 & 0.616 & 0.616 \\
faceID-VGGFace2 - autoenc-VGGFace2 & 0.031 & \([0.009, 0.052]\) & 0.011 & 2.816 & 0.00486 & 0.00638 \\
faceID-VGGFace2 - faceID-BFM & 0.045 & \([0.023, 0.066]\) & 0.011 & 4.053 & \(5.06 \times 10^{-5}\) & \(8.18 \times 10^{-5}\) \\
faceID-VGGFace2 - autoenc-BFM & 0.135 & \([0.114, 0.157]\) & 0.011 & 12.296 & \(1.04 \times 10^{-34}\) & \(2.72 \times 10^{-34}\) \\
invRend-BFM - autoenc-VGGFace2 & 0.025 & \([0.004, 0.047]\) & 0.011 & 2.314 & 0.0207 & 0.0255 \\
invRend-BFM - faceID-BFM & 0.039 & \([0.017, 0.061]\) & 0.011 & 3.551 & \(3.84 \times 10^{-4}\) & \(5.38 \times 10^{-4}\) \\
invRend-BFM - autoenc-BFM & 0.130 & \([0.108, 0.151]\) & 0.011 & 11.794 & \(4.51 \times 10^{-32}\) & \(1.05 \times 10^{-31}\) \\
autoenc-VGGFace2 - faceID-BFM & 0.014 & \([-0.008, 0.035]\) & 0.011 & 1.237 & 0.216 & 0.239 \\
autoenc-VGGFace2 - autoenc-BFM & 0.104 & \([0.083, 0.126]\) & 0.011 & 9.480 & \(2.63 \times 10^{-21}\) & \(5.52 \times 10^{-21}\) \\
faceID-BFM - autoenc-BFM & 0.091 & \([0.069, 0.112]\) & 0.011 & 8.243 & \(1.71 \times 10^{-16}\) & \(3.26 \times 10^{-16}\) \\

\end{longtable}

\end{small}